\documentclass[aps,pra,twocolumn]{revtex4-1}
\usepackage{amsmath,amsthm,amsfonts,amssymb}
\usepackage{xcolor,bm}
\usepackage{mathrsfs}
\usepackage{graphicx}
\usepackage{dsfont}
\usepackage{hyperref}
\usepackage{braket}
\usepackage{todonotes}
\usepackage{tikz}
\usepackage{natbib}
\usepackage{cancel}
\usetikzlibrary{backgrounds,calc,decorations.markings,hobby}
\usepackage{algorithm, algpseudocode}
\usepackage{algorithmicx}

\newtheorem*{remark*}{Remark}

\begin{document}
\title{The matrix permanent and determinant from a spin system}

\author{Abhijeet Alase}
\affiliation{Quantum Science Group, The University of Sydney, NSW 2006,
Australia}

\author{Owen Doty}
\affiliation{Institute for Quantum Science and Technology and Department of
Physics and Astronomy, University of Calgary, Calgary, Alberta T2N 1N4, Canada}

\author{David L. Feder}
\affiliation{Institute for Quantum Science and Technology and Department of
Physics and Astronomy, University of Calgary, Calgary, Alberta T2N 1N4, Canada}

\begin{abstract}

In contrast to the determinant, no algorithm is known for the 
exact determination of the permanent of a square matrix that runs in time 
polynomial in its dimension. Consequently, non-interacting fermions are 
classically efficiently simulatable while non-interacting bosons are not, 
underpinning quantum supremacy arguments for sampling the output distribution 
of photon interferometer arrays. This work introduces a graph-theoretic 
framework that bridges both the determinant and permanent. The only non-zero 
eigenvalues of a sparse non-Hermitian operator $\breve{M}$ for $n$ spin-$1/2$ 
particles are the $n$th roots of the permanent or determinant of an $n\times n$ 
matrix $M$, interpreting basis states as bosonic or fermionic occupation 
states, respectively. This operator can be used to design a simple and 
straightforward method for the classical determination of the permanent that 
matches the efficiency of the best-known algorithm. Gauss-Jordan elimination 
for the determinant of $M$ is then equivalent to the successive removal of the 
generalized zero eigenspace of the fermionic $\breve{M}$, equivalent to the 
deletion of some nodes and reweighting of the remaining edges in the graph such 
that only $n$ nodes survive after the last step. In the bosonic case, the 
successive removal of generalized zero eigenspaces for $\breve{M}$ is also 
equivalent to node deletion, but new edges are added during this process, which 
gives rise to the higher complexity of computing the permanent. Our analysis 
may point the way to new strategies for classical and quantum evaluation of the 
permanent.

\end{abstract}

\maketitle

\section{Introduction}
\label{sec:introduction}

The permanent of a square matrix $M$ of dimension $n$ is the symmetric analogue 
of 
the usual determinant, but where the signatures of the permutations (the signs 
appearing in the expansion of the function) are ignored. This quantity appears
in a wide variety of applications in pure mathematics and in physics, among
other disciplines. For example, the permanent enumerates the number of 
perfect matchings of a bipartite graph, which has applications in 
combinatorics~\cite{Bjorklund2009}, chemistry~\cite{Cvetkovic1972}, and
physics~\cite{Heilmann1972}. The permanent arises in the identification of
multiple targets~\cite{Schaub2017}, with applications to defense. In the 
context of quantum computation and 
information, the permanent is central to calculating matrix elements in linear 
optics for many-photon systems~\cite{Scheel2008,Aaronson2011a,Aaronson2011b}, 
and for determining the entanglement of various permutation-invariant quantum 
states~\cite{Wei2010}.

Despite the fact that both the permanent and the determinant yield the same 
exponential number of terms, $n!\sim\sqrt{2\pi n}(n/e)^n$ for large $n$, the 
determinant is efficiently 
computable classically, i.e.\ scales as a polynomial in $n$. The well-known 
Gaussian elimination approach scales as $O(n^3)$, and the fastest current
algorithm scales as $O(n^{2.373})$~\cite{Fisikopoulos2016}. 
In contrast, determining the permanent of a general matrix is 
\#P-hard, and that of a $(0,1)$ matrix is 
\#P-complete~\cite{Valiant1979a,Valiant1979b,Ben-Dor1993}.
The discovery of a classically efficient algorithm for the permanent would have 
profound consequences for the theory of computation, including 
$\mbox{P}=\mbox{PP}$~\cite{Toda1991}, an even stronger statement than the 
famous $\mbox{P}=\mbox{NP}$ conjecture. The runtime of the fastest known algorithm, 
namely Ryser's algorithm, scales as $O(n2^n)$~\cite{Glynn2010,Niu2020}. That said, the 
permanent $P_n$ of matrices with non-negative entries or with vanishing mean 
can be {\it approximated} in polynomial time $\mbox{poly}(n,1/\epsilon)$ using 
randomized algorithms~\cite{Jerrum2004,Eldar2018}, up to additive error 
$\epsilon P_n$, for arbitrary $\epsilon>0$; likewise for positive semidefinite 
matrices~\cite{Barvinok1997,Gurvits2002,Gurvits2005}. 

The \#P-hardness of computing the permanent was recast in the framework of 
linear optics~\cite{Aaronson2011a}, which motivated the realization that quantum
devices will always outperform classical algorithms in sampling the output
distribution of photons emerging from an optical interferometer apparatus, the
so-called Boson Sampling problem~\cite{Aaronson2011b}. Numerous Boson Sampling
experiments have been conducted since then; 
Refs.~\onlinecite{Arute2019,Zhong2020,Madsen2022} provide some recent examples.
In contrast, the ease of calculating the determinant implies that 
non-interacting fermions are efficiently simulatable on a classical 
computer~\cite{Knill2001,Terhal2002,Valiant2002,Divincenzo2005,Jozsa2008}.

It was recently shown that the permanent of the matrix $M$ can be computed as 
the determinant of a family of matrices $\breve{M}$ of minimum dimension 
$2^n-1$~\cite{Grenet2012,Burgisser2016,Landsberg2016}. These matrices define
the adjacency of a directed $n$-dimensional hypercube graph, whose edge weights 
correspond to elements of the matrix of interest, and with the first and last 
vertices sharing the same label to form a cycle. It was subsequently noted that 
these graphs encode an algebraic branching program~\cite{Huttenhain2016}: the 
product of edge weights on each of the $n!$ possible branches corresponds to a 
term in the expansion of the permanent.

The present work builds on the above construction by identifying a key feature: 
the structure of the matrix $\breve{M}$ coincides with the dynamics of $n$ 
spin-$1/2$ particles governed by a non-Hermitian operator. If the permanent 
of $M$ is non-zero, then the only non-zero eigenvalues of $\breve{M}$ are the 
$n$th roots of the permanent; alternatively, $\breve{M}^n$ diagonalizes into 
$n$ blocks labeled by the total spin, each of which has the permanent as the 
only non-zero eigenvalue. Thus, the $n$-fold product of $\breve{M}$ on a 
fiducial state such as $|0^{\otimes n}\rangle$ immediately yields 
$P_n|0^{\otimes n}\rangle$. The
$n$-sparsity of $\breve{M}$ ensures that this can be effected on a classical 
computer with $n2^n$ arithmetic operations, matching the performance of Ryser's algorithm.

Interpreting the basis states as bosonic occupation states
yields the standard expression for the permanent in terms of products of 
(hard-core) bosonic operators. Interpreting these instead as fermionic 
occupation states immediately yields the determinant, with signed edge weights 
in the graph. If $M$ is a full-rank matrix, then Gaussian elimination for the 
calculation of the determinant corresponds to successively projecting out the 
generalized zero eigenvectors of $\breve{M}$, so that after $n$ iterations the 
initial rank-deficient matrix of dimension $2^n-1$ is reduced to an 
$n$-dimensional full-rank matrix. From the perspective of the algebraic 
branching program, each iteration deletes vertices and the edges incident to 
them, and reweights the remaining edges, until only one path remains in the 
cycle. This approach uncovers another close connection between fermions and the 
determinant on the one hand, and between bosons and the permanent on the other. 

This paper is organized as follows. The permanent and determinant are reviewed
in Sec.~\ref{sec:review}, and an example is provided for the representation of
the permanent as an algebraic branching program. Sec.~\ref{sec:spinmodel}
introduces the spin model that maps the problem of computing the permanent of 
an $n\times n$ matrix $M$ to the problem of computing the eigenvalues of a
$2^n\times 2^n$ matrix $\breve{M}$, and provides a classical algorithm for
computing the permanent that matches the best current methods. The spin model 
is expressed in terms of non-interacting fermions and hard-core bosons in 
Sec.~\ref{sec:mapping}. In Sec.~\ref{sec:rowreduce}, we discuss the connection 
between Gaussian elimination, generalized zero eigenspaces of $\breve{M}$ and 
its visualization on the associated graph. The prospects for the development of 
a quantum algorithm for computing the permanent based on our approach are 
discussed in Sec.~\ref{sec:prospects}.

\section{Review}
\label{sec:review}

\subsection{Permanent and determinant}
\label{subsec:permderm}

Consider the $n\times n$ matrix $M$, defined as
\begin{equation}
M=\begin{pmatrix}
w_{0,0} & w_{0,1} & \cdots & w_{0,n-1}\cr
w_{1,0} & w_{1,1} & \cdots & w_{1,n-1}\cr
\vdots & \vdots & \ddots & \vdots\cr
w_{n-1,0} & w_{n-1,1} & \cdots & w_{n-1,n-1}\cr
\end{pmatrix}.
\end{equation}
The determinant and permanent of $M$ are respectively defined as
\begin{eqnarray}
D_n&=&\left|M\right|=\det(M)\equiv\sum_{\sigma\in S_n}\left(\mbox{sgn}(\sigma)
\prod_{i=0}^{n-1}w_{i,\sigma_i}\right);\nonumber \\
P_n&=&\left|M\right|_P=\mbox{perm}(M)\equiv\sum_{\sigma\in S_n}\left(
\prod_{i=0}^{n-1}w_{i,\sigma_i}\right),\nonumber
\end{eqnarray}
where $S_n$ is the symmetric group on the list $\{0,1,2,\ldots,n-1\}$, $\sigma$ 
is a function that reorders this list (effects a permutation of the elements),
$\sigma_i$ is the $i$th entry of the list after permutation, and
$\mbox{sgn}(\sigma)=(-1)^{N(\sigma)}$ is the signature of the permutation,
where $N(\sigma)$ is the number of inversions needed. While the expansion of
the determinant and permanent includes the same $n!$ terms, the signs appearing
in the determinant allow for its efficient evaluation.

While exceedingly simple, the $n=3$ case is illustrative and will be revisited 
throughout this work. The determinant is explicitly written
\begin{eqnarray}
|M|&=&w_{0,0}\left(w_{1,1}w_{2,2}-w_{1,2}w_{2,1}\right)\nonumber \\
&-&w_{0,1}\left(w_{1,0}w_{2,2}-w_{1,2}w_{2,0}\right)\nonumber \\
&+&w_{0,2}\left(w_{1,0}w_{2,1}-w_{1,1}w_{2,0}\right).
\label{eq:detn=3}
\end{eqnarray}
The Gaussian elimination algorithm uses pivoting to reduce the matrix to 
row echelon form (i.e.\ an upper triangular matrix), so that the determinant is
the product of the diagonal elements. For reasons that will become clear in
Sec.~\ref{sec:mapping}, consider instead a reduction to a lower triangular
matrix. The first reduction yields
\begin{equation}
|M|=\left|\begin{matrix}
w_{0,0}' & w_{0,1}' & 0\cr
w_{1,0}' & w_{1,1}' & 0\cr
w_{2,0} & w_{2,1} & w_{2,2}\cr\end{matrix}\right|,
\end{equation}
where
\begin{eqnarray}
w_{0,0}'&=&w_{0,0}-\frac{w_{0,2}w_{1,0}}{w_{1,2}};\quad 
w_{0,1}'=w_{0,1}-\frac{w_{0,2}w_{1,1}}{w_{1,2}};\label{eq:primed} \\
w_{1,0}'&=&w_{1,0}-\frac{w_{1,2}w_{2,0}}{w_{2,2}};\quad
w_{1,1}'=w_{1,1}-\frac{w_{1,2}w_{2,1}}{w_{2,2}}.\label{eq:primed2}
\end{eqnarray}
The second and last reduction yields
\begin{equation}
|M|=\left|\begin{matrix}
w_{0,0}'' & 0 & 0\cr
w_{1,0}' & w_{1,1}' & 0\cr
w_{2,0} & w_{2,1} & w_{2,2}\cr\end{matrix}\right|,
\label{eq:w22pp}
\end{equation}
where
\begin{equation}
w_{0,0}''=w_{0,0}'-\frac{w_{0,1}'w_{1,0}'}{w_{1,1}'}.
\label{eq:w00pp}
\end{equation}
The determinant is then
\begin{eqnarray}
|M|&=&w_{0,0}''w_{1,1}'w_{2,2}
=\left(w_{0,0}'w_{1,1}'-w_{0,1}'w_{1,0}'\right)w_{2,2}\nonumber \\
&=&\left(w_{0,0}-\frac{w_{0,2}w_{1,0}}{w_{1,2}}\right)
\left(w_{1,1}-\frac{w_{1,2}w_{2,1}}{w_{2,2}}\right)\nonumber \\
&-&\left(w_{0,1}-\frac{w_{0,2}w_{1,1}}{w_{1,2}}\right)
\left(w_{1,0}-\frac{w_{1,2}w_{2,0}}{w_{2,2}}\right)w_{2,2}.\hphantom{a}
\label{eq:upper}
\end{eqnarray}
While there are eight terms in the expansion, the signs the two cross terms
$\left(-\frac{w_{0,2}w_{1,0}}{w_{1,2}}\right)\left(w_{1,1}\right)w_{2,2}$
and $-\left(-\frac{w_{0,2}w_{1,1}}{w_{1,2}}\right)\left(w_{1,0}\right)w_{2,2}$
cancel, leaving 6 unique terms in the expansion.

The sign structure of the determinant guarantees that these cancellations occur 
for all values of $n$, which ensures that Gaussian elimination is 
classically efficient. For the evaluation of the permanent, one cannot follow 
the same procedure as above by simply eliminating all signs, because the cross 
terms arising from expanding the final product [for example in 
Eq.~(\ref{eq:upper})] will now add instead of cancelling. Our analysis
in Sec.~\ref{subsec:rowreducebosons} provides insight into why this is the case.

\subsection{Permanent as an algebraic branching program}
\label{subsec:ABP}

\begin{figure}[t]
\begin{tikzpicture}[decoration={
    markings,
    mark=at position 0.35 with {\arrow{>}}},
		 scale=0.98,every node/.style={scale=0.98}
]

\draw[fill,red] (1.55, 1.0) arc(0:360:0.05) -- cycle;
\draw[fill] (0.55, 0.0) arc(0:360:0.05) -- cycle;
\draw[fill] (1.55, 0.0) arc(0:360:0.05) -- cycle;
\draw[fill] (2.55, 0.0) arc(0:360:0.05) -- cycle;
\draw[fill] (0.55, -1.0) arc(0:360:0.05) -- cycle;
\draw[fill] (1.55, -1.0) arc(0:360:0.05) -- cycle;
\draw[fill] (2.55, -1.0) arc(0:360:0.05) -- cycle;
\draw[fill,red] (1.55, -2.0) arc(0:360:0.05) -- cycle;

\draw[thick,black,postaction={decorate}] (1.5,1) -- (0.5,0)
  node [midway,left] {\footnotesize $w_{\scalebox{.8}{$\scriptstyle 0,0$}}$};
\draw[thick,black,postaction={decorate}] (1.5,1) -- (1.5,0)
  node [pos=0.7,right] {\footnotesize $w_{\scalebox{.8}{$\scriptstyle 1,0$}}$};
\draw[thick,black,postaction={decorate}] (1.5,1) -- (2.5,0)
  node [midway,right] {\footnotesize $w_{\scalebox{.8}{$\scriptstyle 2,0$}}$};
\draw[thick,black,postaction={decorate}] (0.5,0) -- (0.5,-1)
  node [midway,left] {\footnotesize $w_{\scalebox{.8}{$\scriptstyle 1,1$}}$};
\draw[thick,black,postaction={decorate}] (0.5,0) -- (1.5,-1)
  node [pos=0.3,above] {\footnotesize $w_{\scalebox{.8}{$\scriptstyle 2,1$}}$};
\draw[thick,black,postaction={decorate}] (1.5,0) -- (0.5,-1)
  node [near start,above] {\footnotesize $w_{\scalebox{.8}{$\scriptstyle 0,1$}}$};
\draw[thick,black,postaction={decorate}] (1.5,0) -- (2.5,-1)
  node [pos=0.3,above] {\footnotesize $w_{\scalebox{.8}{$\scriptstyle 2,1$}}$};
\draw[thick,black,postaction={decorate}] (2.5,0) -- (2.5,-1)
  node [midway,right] {\footnotesize $w_{\scalebox{.8}{$\scriptstyle 1,1$}}$};
\draw[thick,black,postaction={decorate}] (2.5,0) -- (1.5,-1)
  node [near start,above] {\footnotesize $w_{\scalebox{.8}{$\scriptstyle 0,1$}}$};
\draw[thick,black,postaction={decorate}] (0.5,-1) -- (1.5,-2)
  node [midway,left] {\footnotesize $w_{\scalebox{.8}{$\scriptstyle 2,2$}}$};
\draw[thick,black,postaction={decorate}] (1.5,-1) -- (1.5,-2)
  node [pos=0.2,right] {\footnotesize $w_{\scalebox{.8}{$\scriptstyle 1,2$}}$};
\draw[thick,black,postaction={decorate}] (2.5,-1) -- (1.5,-2)
  node [midway,right] {\footnotesize $w_{\scalebox{.8}{$\scriptstyle 0,2$}}$};


\draw[thick,black] (3.25,0.5) node {$=$};

\draw[fill,red] (5.05, 1.0) arc(0:360:0.05) -- cycle;
\draw[fill] (4.05, 0.0) arc(0:360:0.05) -- cycle;
\draw[fill] (4.05, -1.0) arc(0:360:0.05) -- cycle;
\draw[fill] (5.05, -1.0) arc(0:360:0.05) -- cycle;
\draw[fill,red] (5.05, -2.0) arc(0:360:0.05) -- cycle;


\draw[thick,black,postaction={decorate}] (5,1) -- (4,0)
  node [midway,left] {\footnotesize $w_{\scalebox{.8}{$\scriptstyle 0,0$}}$};
\draw[thick,black,postaction={decorate}] (4,0) -- (4,-1)
  node [midway,left] {\footnotesize $w_{\scalebox{.8}{$\scriptstyle 1,1$}}$};
\draw[thick,black,postaction={decorate}] (4,0) -- (5,-1)
  node [pos=0.4,above] {\footnotesize $w_{\scalebox{.8}{$\scriptstyle 2,1$}}$};
\draw[thick,black,postaction={decorate}] (4,-1) -- (5,-2)
  node [midway,left] {\footnotesize $w_{\scalebox{.8}{$\scriptstyle 2,2$}}$};
\draw[thick,black,postaction={decorate}] (5,-1) -- (5,-2)
  node [midway,right] {\footnotesize $w_{\scalebox{.8}{$\scriptstyle 1,2$}}$};

\draw[thick,black] (5.0,0.5) node {$+$};

\draw[fill,red] (6.05, 2.0) arc(0:360:0.05) -- cycle;
\draw[fill] (6.05, 1.0) arc(0:360:0.05) -- cycle;
\draw[fill] (5.05, 0.0) arc(0:360:0.05) -- cycle;
\draw[fill] (7.05, 0.0) arc(0:360:0.05) -- cycle;
\draw[fill,red] (6.05, -1.0) arc(0:360:0.05) -- cycle;

\draw[thick,black,postaction={decorate}] (6,2) -- (6,1)
  node [midway,right] {\footnotesize $w_{\scalebox{.8}{$\scriptstyle 1,0$}}$};
\draw[thick,black,postaction={decorate}] (6,1) -- (5,0)
  node [pos=0.6,above] {\footnotesize $w_{\scalebox{.8}{$\scriptstyle 0,1$}}$};
\draw[thick,black,postaction={decorate}] (6,1) -- (7,0)
  node [pos=0.6,above] {\footnotesize $w_{\scalebox{.8}{$\scriptstyle 2,1$}}$};
\draw[thick,black,postaction={decorate}] (5,0) -- (6,-1)
  node [midway,left] {\footnotesize $w_{\scalebox{.8}{$\scriptstyle 2,2$}}$};
\draw[thick,black,postaction={decorate}] (7,0) -- (6,-1)
  node [midway,right] {\footnotesize $w_{\scalebox{.8}{$\scriptstyle 0,2$}}$};


\draw[thick,black] (7.0,0.5) node {$+$};

\draw[fill,red] (7.05, 1.0) arc(0:360:0.05) -- cycle;
\draw[fill] (8.05, 0.0) arc(0:360:0.05) -- cycle;
\draw[fill] (7.05, -1.0) arc(0:360:0.05) -- cycle;
\draw[fill] (8.05, -1.0) arc(0:360:0.05) -- cycle;
\draw[fill,red] (7.05, -2.0) arc(0:360:0.05) -- cycle;

\draw[thick,black,postaction={decorate}] (7,1) -- (8,0)
  node [midway,right] {\footnotesize $w_{\scalebox{.8}{$\scriptstyle 2,0$}}$};
\draw[thick,black,postaction={decorate}] (8,0) -- (8,-1)
  node [midway,right] {\footnotesize $w_{\scalebox{.8}{$\scriptstyle 1,1$}}$};
\draw[thick,black,postaction={decorate}] (8,0) -- (7,-1)
  node [near start,above] {\footnotesize $w_{\scalebox{.8}{$\scriptstyle 0,1$}}$};
\draw[thick,black,postaction={decorate}] (7,-1) -- (7,-2)
  node [pos=0.2,right] {\footnotesize $w_{\scalebox{.8}{$\scriptstyle 1,2$}}$};
\draw[thick,black,postaction={decorate}] (8,-1) -- (7,-2)
  node [midway,right] {\footnotesize $w_{\scalebox{.8}{$\scriptstyle 0,2$}}$};


\end{tikzpicture}
\caption{(Color online) Illustration of the algebraic branching program for the 
evaluation of the permanent $\left|M\right|_P$ for $n=3$.}
\label{fig:ABP}
\end{figure}
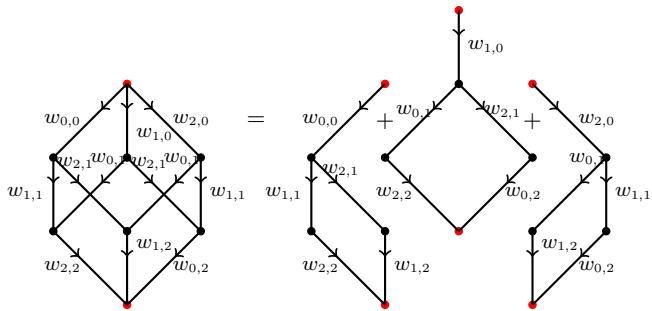

Building on the work of Grenet and 
others~\cite{Grenet2012,Burgisser2016,Landsberg2016}, H\" uttenhain and
Ikenmeyer~\cite{Huttenhain2016} noted that the matrix permanent for $n=3$ can 
be expressed as a binary algebraic branching program. The $n!$ terms correspond 
to branches, or routes, traversing between antipodes of the $n$-dimensional 
hypercube, such that the product of edge weights for each branch corresponds to 
a term in the expansion of the permanent. Fig.~\ref{fig:ABP} illustrates the 
idea for the $n=3$ case, where the three main branches from the top to bottom 
vertices (labeled in red) are explicitly shown. The edge weights are chosen so 
that their products for each branch correspond to a term in the permanent; 
c.f.\ Eq.~(\ref{eq:detn=3}) with signs removed. The branching program is the 
analog of the expansion of the determinant by matrix minors.

\section{Spin model}
\label{sec:spinmodel}

\subsection{Definition and structure}
\label{subsec:definition}

The binary algebraic branching program for the $3\times 3$ 
permanent~\cite{Huttenhain2016} suggests a general construction for arbitrary
$n$. Suppose one has a system of spin-$1/2$ particles, located on sites 
$j=0,1,\ldots,n-1$. Each particle can access states $|0\rangle$ and 
$|1\rangle$, corresponding to spin down and spin up respectively. The spin
model that is the central focus of the current work is defined by the operator 
\begin{equation}
\tilde{M}=\sum_{\bf i}\sum_{j=0}^{n-1}w_{h({\bf i}),j}\sigma^+_j
|{\bf i}\rangle\langle{\bf i}|+\prod_j\sigma^-_j,
\label{eq:Ham}
\end{equation}
where $\sigma^+_i=|1\rangle\langle 0|_i$ and $\sigma^-_i=|0\rangle\langle 1|_i$
are site-dependent raising and lowering operators. The first sum is over all
$n$-bit strings ${\bf i}$ so that a complete and orthonormal basis of $n$-spin 
states with dimension $2^n$ is represented by the unit vectors $|{\bf i}\rangle
=|\{0,1\}\rangle^{\otimes n}$. The Hamming weight of the bitstring is denoted 
by $h({\bf i})$, coinciding with the total $n$-particle spin. Evidently the 
last term in Eq.~(\ref{eq:Ham}) is equivalent to 
$|{\bf 0}\rangle\langle{\bf 1}|$.

\begin{figure}[t]
\begin{tikzpicture}[decoration={
    markings,
    mark=at position 0.35 with {\arrow{>}}},
		scale=0.75,every node/.style={scale=0.75}
]


\draw[fill] (2.55, 2.0) arc(0:360:0.05) -- cycle;
\draw[fill] (0.55, 0.0) arc(0:360:0.05) -- cycle;
\draw[fill] (2.55, 0.0) arc(0:360:0.05) -- cycle;
\draw[fill] (4.55, 0.0) arc(0:360:0.05) -- cycle;
\draw[fill] (0.55, -2.0) arc(0:360:0.05) -- cycle;
\draw[fill] (2.55, -2.0) arc(0:360:0.05) -- cycle;
\draw[fill] (4.55, -2.0) arc(0:360:0.05) -- cycle;
\draw[fill] (2.55, -4.0) arc(0:360:0.05) -- cycle;

\draw[thick,black,postaction={decorate}] (2.5,2) -- (0.5,0)
  node [midway,left] {\small $w_{0,0}$};
\draw[thick,black,postaction={decorate}] (2.5,2) -- (2.5,0)
  node [pos=0.7,right] {\small $w_{0,1}$};
\draw[thick,black,postaction={decorate}] (2.5,2) -- (4.5,0)
  node [midway,right] {\small $w_{0,2}$};
\draw[thick,black,postaction={decorate}] (0.5,0) -- (0.5,-2)
  node [midway,left] {\small $w_{1,1}$};
\draw[thick,black,postaction={decorate}] (0.5,0) -- (2.5,-2)
  node [pos=0.3,above] {\small $w_{1,2}$};
\draw[thick,black,postaction={decorate}] (2.5,0) -- (0.5,-2)
  node [near start,above] {\small $w_{1,0}$};
\draw[thick,black,postaction={decorate}] (2.5,0) -- (4.5,-2)
  node [pos=0.3,above] {\small $w_{1,2}$};
\draw[thick,black,postaction={decorate}] (4.5,0) -- (4.5,-2)
  node [midway,right] {\small $w_{1,1}$};
\draw[thick,black,postaction={decorate}] (4.5,0) -- (2.5,-2)
  node [near start,above] {\small $w_{1,0}$};
\draw[thick,black,postaction={decorate}] (0.5,-2) -- (2.5,-4)
  node [midway,left] {\small $w_{2,2}$};
\draw[thick,black,postaction={decorate}] (2.5,-2) -- (2.5,-4)
  node [pos=0.2,right] {\small $w_{2,1}$};
\draw[thick,black,postaction={decorate}] (4.5,-2) -- (2.5,-4)
  node [near start,right] {\small $w_{2,0}$};

\draw[thick,black] (2.5,2.2) node {$|000\rangle$};
\draw[thick,black] (0.05,0.0) node {$|100\rangle$};
\draw[thick,black] (2.9,-0.0) node {$|010\rangle$};
\draw[thick,black] (4.95,0.0) node {$|001\rangle$};
\draw[thick,black] (0.05,-2.0) node {$|110\rangle$};
\draw[thick,black] (2.9,-2.0) node {$|101\rangle$};
\draw[thick,black] (4.95,-2.0) node {$|011\rangle$};
\draw[thick,black] (2.5,-4.2) node {$|111\rangle$};

\draw[thick,black,postaction={decorate}] (2.5, -4.0) arc(-60:60:3.45);

\draw[] (2.5,-4.8) node {(a)};


\draw[fill] (8.55, 2.0) arc(0:360:0.05) -- cycle;
\draw[fill] (6.55, 0.0) arc(0:360:0.05) -- cycle;
\draw[fill] (8.55, 0.0) arc(0:360:0.05) -- cycle;
\draw[fill] (10.55, 0.0) arc(0:360:0.05) -- cycle;
\draw[fill] (6.55, -2.0) arc(0:360:0.05) -- cycle;
\draw[fill] (8.55, -2.0) arc(0:360:0.05) -- cycle;
\draw[fill] (10.55, -2.0) arc(0:360:0.05) -- cycle;
\draw[fill] (8.55, -4.0) arc(0:360:0.05) -- cycle;

\draw[thick,black,postaction={decorate}] (8.5,2) -- (6.5,0)
  node [midway,left] {\small $w_{0,0}$};
\draw[thick,black,postaction={decorate}] (8.5,2) -- (8.5,0)
  node [pos=0.7,right] {\small $w_{0,1}$};
\draw[thick,black,postaction={decorate}] (8.5,2) -- (10.5,0)
  node [midway,right] {\small $w_{0,2}$};
\draw[thick,black,postaction={decorate}] (6.5,0) -- (6.5,-2)
  node [midway,left] {\small $w_{1,1}$};
\draw[thick,black,postaction={decorate}] (6.5,0) -- (8.5,-2)
  node [pos=0.3,above] {\small $w_{1,2}$};
\draw[thick,black,postaction={decorate}] (8.5,0) -- (6.5,-2)
  node [near start,above] {\small $w_{1,0}$};
\draw[thick,black,postaction={decorate}] (8.5,0) -- (10.5,-2)
  node [pos=0.3,above] {\small $w_{1,2}$};
	\draw[thick,black,postaction={decorate}] (10.5,0) -- (10.5,-2)
  node [midway,right] {\small $w_{1,1}$};
\draw[thick,black,postaction={decorate}] (10.5,0) -- (8.5,-2)
  node [near start,above] {\small $w_{1,0}$};
\draw[thick,black,postaction={decorate}] (6.5,-2) -- (8.5,-4)
  node [midway,left] {\small $w_{2,2}$};
\draw[thick,black,postaction={decorate}] (8.5,-2) -- (8.5,-4)
  node [pos=0.2,right] {\small $w_{2,1}$};
\draw[thick,black,postaction={decorate}] (10.5,-2) -- (8.5,-4)
  node [midway,right] {\small $w_{2,0}$};

\draw[thick,black] (8.5,2.2) node {$|000\rangle$};
\draw[thick,black] (6.05,0.0) node {$|100\rangle$};
\draw[thick,black] (8.9,-0.0) node {$|010\rangle$};
\draw[thick,black] (10.95,0.0) node {$|001\rangle$};
\draw[thick,black] (6.05,-2.0) node {$|110\rangle$};
\draw[thick,black] (8.9,-2.0) node {$|101\rangle$};
\draw[thick,black] (10.95,-2.0) node {$|011\rangle$};
\draw[thick,black] (8.5,-4.2) node {$|000\rangle$};

\draw[] (8.5,-4.8) node {(b)};

\end{tikzpicture}
\caption{Depictions of the spin model $\tilde{M}$ (a) and its alternative
description $\breve{M}$ (b), for $n=3$.}
\label{fig:Hamn=3}
\end{figure}
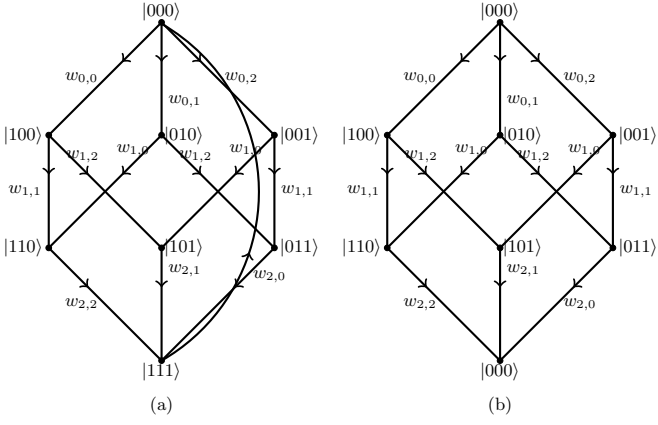

The operator $\tilde{M}$ defined by Eq.~(\ref{eq:Ham}) corresponds to 
the adjacency matrix for a weighted directed graph that effects transitions 
from the $|{\bf 0}\rangle$ state to the $|{\bf 1}\rangle$ state via all 
possible 
single-spin raising operations, and then back to $|{\bf 0}\rangle$ again to 
complete one cycle. The transition amplitudes are indexed by two integers: the 
total Hamming weight of the initial state and the target site. With 
$\sigma^+|1\rangle=0$, the second index can never be repeated as the value of 
first index increases; thus, the first term in $\tilde{M}$ encodes all possible 
transitions from $|{\bf 0}\rangle$ to $|{\bf 1}\rangle$ without repetitions. 
Fig.~\ref{fig:Hamn=3}(a) depicts $\tilde{M}$ for $n=3$, and includes the 
vertex / state labelings for clarity. The orientation is chosen so that each 
horizontal layer of the hypercube contains vertices labeled by bitstrings with 
the same Hamming weight $h$.

As discussed in detail in what follows, it is convenient to define an
alternate encoding of the cyclic behavior of $\tilde{M}$ by eliminating
the transition $|{\bf 0}\rangle\langle{\bf 1}|$, and instead directly 
transition from states with Hamming weight $n-1$ to the state 
$|{\bf 0}\rangle$. The associated operator is
\begin{equation}
\breve{M}={\sum_{\bf i}}'\sum_{j=0}^{n-1}w_{h({\bf i}),j}\sigma^+_j
|{\bf i}\rangle\langle{\bf i}|
+\sum_{j=0}^{n-1}w_{n-1,j}|{\bf 0}\rangle\langle{\bf 1}|\sigma_j^+,
\label{eq:Hamalt}
\end{equation}
where the prime on the first term denotes that the sum is over all bitstrings 
but not including those with Hamming weight $h({\bf i})=n-1$. In this case, the
basis state $|{\bf 1}\rangle$ is never occupied, and the Hilbert space 
dimension is reduced to $2^n-1$. This alternate operator is depicted in 
Fig.~\ref{fig:Hamn=3}(b).

Consider next the $(n+1)$th ($n$th) power of $\tilde{M}$ ($\breve{M}$), which
will be of central importance in what follows. The derivation is provided in 
Appendix~\ref{app:block}, and the result for $\tilde{M}^{n+1}$ is given in
Eq.~(\ref{eq:Mnapp}):
\begin{eqnarray}
\tilde{M}^{n+1}&=&\sum_{j_0,\ldots,j_{n-1}}\left[
\left(w_{0,j_0}\sigma^+_{j_0}\right)\cdots\left(w_{n-1,j_{n-1}}
\sigma^+_{j_{n-1}}\right)|{\bf 0}\rangle\langle{\bf 1}|\right.\nonumber \\
&+&\left(w_{0,j_0}\sigma^+_{j_0}\right)\cdots\left(w_{n-2,j_{n-2}}
\sigma^+_{j_{n-2}}\right)|{\bf 0}\rangle\langle{\bf 1}|\nonumber \\
&&\quad\times\left(w_{n-1,j_{n-1}}\sigma^+_{j_{n-1}}\right)\nonumber \\
&+&\ldots+\left.|{\bf 0}\rangle\langle{\bf 1}|\left(w_{0,j_0}\sigma^+_{j_0}
\right)\cdots\left(w_{n-1,j_{n-1}}\sigma^+_{j_{n-1}}\right)\right].\nonumber \\
\label{eq:Mn}
\end{eqnarray}
The expression for $(\breve{M})^n$ is identical save for the leading term in 
Eq.~(\ref{eq:Mn}). 

The above expression can be seen to be of block diagonal form as follows. Each
term in the expansion above is defined by the operators 
$\prod_{r=0}^{m-1}\sigma^+_{j_r}|{\bf 0}\rangle\langle{\bf 1}|
\prod_{s=m}^{n-1}\sigma^+_{j_s}$, and is labeled by the index $m=0,1,\ldots,n$.
For $m=0$, the operator is only $|{\bf 0}\rangle\langle{\bf 0}|$, i.e.\ a block 
of dimension 1 defined by a single basis state with zero Hamming weight. The 
$m=1$ case includes all operators of the form
\begin{equation}
\sigma^+_{j_0}|{\bf 0}\rangle\langle{\bf 1}|\prod_{s=1}^{n-1}\sigma^+_{j_s}
=\sigma^+_{j_0}|{\bf 0}\rangle\langle{\bf 1}|\prod_{s=0}^{n-1}\sigma^+_{j_s}
\sigma^-_{j_0}
=\sigma^+_{j_0}|{\bf 0}\rangle\langle{\bf 0}|\sigma^-_{k_0},
\end{equation}
which corresponds to a block spanned by the $n$ basis vectors defined by
$\sigma^+_j|{\bf 0}\rangle$, which are labeled by all bitstrings with unit 
Hamming weight. Evidently, each block is indexed by the Hamming weight (or 
total spin) $m$, and has dimension given by the binomial factor 
$\left({n\atop m}\right)$. It is convenient to express $\tilde{M}^{n+1}$ as the 
direct sum
\begin{equation}
\tilde{M}^{n+1}=\tilde{M}_0\oplus\tilde{M}_1\oplus\cdots\oplus\tilde{M}_n
=\bigoplus_{m=0}^n\tilde{M}_m,
\end{equation}
where $\tilde{M}_m$ corresponds to the block matrix labeled by $m$ and is 
defined as
\begin{equation}
\tilde{M}_m=\tilde{M}^m|{\bf 0}\rangle\langle{\bf 0}|\tilde{M}^{n-m+1},
\label{eq:Mblock}
\end{equation}
which has the form
\begin{eqnarray}
\tilde{M}_m&=&\sum_{j_0,\ldots,j_{n-1}}\left(w_{0,j_0}\sigma^+_{j_0}\right)
\left(w_{1,j_1}\sigma_{j_1}^+\right)\cdots\nonumber \\
&\times&\left(w_{m-1,j_{m-1}}\sigma_{j_{m-1}}^+\right)|{\bf 0}\rangle
\langle{\bf 1}|\left(w_{n-1,j_{n-1}}\sigma^+_{j_{n-1}}\right)\nonumber \\
&\times&\left(w_{n-2,j_{n-2}}\sigma^+_{j_{n-2}}\right)\cdots\left(w_{m,j_m}
\sigma^+_{j_m}\right),
\label{eq:Mblock2}
\end{eqnarray}
as proven as Eq.~(\ref{app:Mm}) in Appendix~\ref{app:block}. Likewise,
\begin{equation}
\breve{M}_m=\breve{M}^m|{\bf 0}\rangle\langle{\bf 0}|\breve{M}^{n-m}.
\label{eq:Mblockprime}
\end{equation}

\subsection{Eigensystem}
\label{subsec:eigensystem}

Now turn to the eigenvalues and eigenvectors of the spin model, defined by
Eq.~(\ref{eq:Ham}) or its alternative expression Eq.~(\ref{eq:Hamalt}). A key
observation is that $|{\bf 0}\rangle$ is an eigenvector of $\tilde{M}^{n+1}$
or $\breve{M}^n$. Consider the action of $\tilde{M}^{n+1}$ on
the state $|{\bf 0}\rangle$, which only involves the $m=0$ block:
\begin{eqnarray}
\tilde{M}^{n+1}|{\bf 0}\rangle&=&
\sum_{j_0,\ldots,j_{n-1}}
|{\bf 0}\rangle\langle{\bf 1}|\left(w_{0,j_0}w_{1,j_1}\cdots w_{n-1,j_{n-1}}
\right)\nonumber \\
&&\quad\times\sigma^+_{j_0}\sigma^+_{j_1}\cdots\sigma^+_{j_{n-1}}
|{\bf 0}\rangle.
\end{eqnarray}
The action of $\sigma^+_{j_0}\sigma^+_{j_1}\cdots\sigma^+_{j_{n-1}}$ defines
all possible $n$ spin-flip paths from the $|{\bf 0}\rangle$ state to 
$|{\bf 1}\rangle$, and each is weighted by the factor 
$w_{0,j_0}w_{1,j_1}\cdots w_{n-1,j_{n-1}}$. This is precisely the algebraic
branching program discussed in Sec.~\ref{subsec:ABP}; thus
\begin{equation}
\tilde{M}^{n+1}|{\bf 0}\rangle=\tilde{M}_0|{\bf 0}\rangle=P_n|{\bf 0}\rangle;
\label{eq:M0}
\end{equation}
the eigenvalue is the permanent of $M$. Likewise, 
$\breve{M}^n|{\bf 0}\rangle=P_n|{\bf 0}\rangle$.

The permanent is also an eigenvalue of every other block of $\tilde{M}^{n+1}$.
Defining the block-$m$ state
\begin{equation}
|\psi_m\rangle=\tilde{M}^m|{\bf 0}\rangle,
\end{equation}
one obtains
\begin{eqnarray}
\tilde{M}_m|\psi_m\rangle&=&\tilde{M}^m|{\bf 0}\rangle\langle{\bf 0}|
\tilde{M}^{n-m+1}\tilde{M}^m|{\bf 0}\rangle\nonumber \\
&=&\tilde{M}^m|{\bf 0}\rangle\langle{\bf 0}|\tilde{M}^{n+1}|{\bf 0}\rangle
=P_n\tilde{M}^m|{\bf 0}\rangle\nonumber \\
&=&P_n|\psi_m\rangle.
\end{eqnarray}
The operator $\tilde{M}^{n+1}$ therefore has $n+1$ degenerate eigenvalues
corresponding to the permanent, with associated eigenvectors 
$|\psi_m\rangle=\tilde{M}^m|{\bf 0}\rangle$. Likewise, the operator
$\breve{M}^n$ has $n$ degenerate eigenvalues $P_n$ and 
associated eigenvectors $\breve{M}^m|{\bf 0}\rangle$.

For the rest of the discussion in this section, we assume that $P_n \ne 0$.
Because $\tilde{M}$ is a cycle, if $\lambda$ is an eigenvalue of 
$\tilde{M}^{n+1}$, then the eigenvalues $\lambda_j$ of $\tilde{M}$ must include 
all $(n+1)$th roots of $\lambda$~(see for example 
Ref.~\onlinecite{Watkins2004}). For the present case $\lambda=P_n$, one
obtains $\lambda_j=P_n^{1/(n+1)}e^{-i2\pi j/(n+1)},\,j=0,1,\ldots,n$; likewise, 
the eigenvalues of $\breve{M}$ are $\left(P_ne^{-i2\pi j}\right)^{1/n},\,
j=0,1,\ldots,n-1$. Given that $\breve{M}$ has degeneracy $n$ and therefore
only requires $n$ powers to return the state $|{\bf 0}\rangle$ to itself, it is 
slightly more convenient to work with $\breve{M}$ in what follows.

The eigenvectors of $\breve{M}$ with eigenvalues corresponding to the $n$th
roots of $P_n$ can be written as
\begin{equation}
|\phi_n(k)\rangle=e^{-i2\pi k/n}\sum_{j=0}^{n-1}e^{i2\pi jk/n}
\left(\frac{\breve{M}}{P_n^{1/n}}\right)^j|0^{\otimes n}\rangle,
\label{eq:eigenvectors}
\end{equation}
where $k,j=0,1,\ldots.n-1$. The corresponding eigenvalues can be found 
directly:
\begin{eqnarray}
&&\breve{M}|\phi_n(k)\rangle=e^{-i2\pi k/n}\sum_{j=0}^{n-1}e^{i2\pi jk/n}
P_n^{1/n}\left(\frac{\breve{M}}{P_n^{1/n}}\right)^{j+1}|0^{\otimes n}\rangle
\nonumber \\
&&\qquad\quad=e^{-i4\pi k/n}P_n^{1/n}\sum_{j=0}^{n-1}e^{i2\pi(j+1)k/n}
\left(\frac{\breve{M}}{P_n^{1/n}}\right)^{j+1}|0^{\otimes n}\rangle\nonumber \\
&&\qquad\quad=e^{-i4\pi k/n}P^{1/n}\sum_{j=1}^ne^{i2\pi jk/n}\left(
\frac{\breve{M}}{P_n^{1/n}}\right)^j|0^{\otimes n}\rangle\nonumber \\
&&\qquad\quad=e^{-i4\pi k/n}P_n^{1/n}\sum_{j=0}^{n-1}e^{i2\pi jk/n}\left(
\frac{\breve{M}}{P_n^{1/n}}\right)^j|0^{\otimes n}\rangle\nonumber \\
&&\qquad\quad+|0^{\otimes n}\rangle-|0^{\otimes n}\rangle\nonumber \\
&&\qquad\quad=e^{-i2\pi k/n}P_n^{1/n}|\phi_n(k)\rangle.
\end{eqnarray}
The eigenvalues are therefore 
\begin{equation}
\lambda_k\left(\breve{M}\right)=e^{-i2\pi k/n}P_n^{1/n}
=\left(e^{-i2\pi k}P_n\right)^{1/n}.
\label{eq:eigenvalues}
\end{equation}
The derivation proceeds analogously for $\tilde{M}$, and one obtains
\begin{equation}
\lambda_k\left(\tilde{M}\right)=\left(e^{-i2\pi k}P_n\right)^{1/(n+1)}.
\label{eq:eigenvaluesb}
\end{equation}
The simplest case corresponds to $k=0$:
\begin{equation}
|\phi_n(0)\rangle=\sum_{j=0}^{n-1}\left(\frac{\breve{M}}{P_n^{1/n}}\right)^j
|0^{\otimes n}\rangle,
\label{eq:eigenvector0}
\end{equation}
with eigenvalue $\lambda_0=P_n^{1/n}$. Consequently, $\pm\lambda_0$ would be the only real eigenvalues
if the elements of $M$ were real and positive. 
Remarkably, Eq.~(\ref{eq:eigenvalues}) and (\ref{eq:eigenvaluesb}) constitute 
the only non-zero eigenvalues of $\breve{M}$ and $\tilde{M}$, respectively.

The periodic nature of $\tilde{M}$ and $\breve{M}$ gives rise to eigenvectors 
that are expanded in a Fourier-like series, much like in a translationally
invariant system. In Eq.~(\ref{eq:eigenvectors}) and Eq.~(\ref{eq:eigenvalues}),
the indices $j$ and $k$ label `position' and `wavevector', respectively. In the 
present case, the position is the index of the block, corresponding to the 
Hamming weight or total spin, while the `wavevector' serves essentially the 
same purpose as in uniform systems: as a canonically conjugate quantum number.
Conceptually, one can consider successive applications of $\tilde{M}$ or 
$\breve{M}$ as moving a walker from `site' $|{\bf 0}\rangle$ to `site'
$\tilde{M}|{\bf 0}\rangle$ or $\breve{M}|{\bf 0}\rangle$, etc., one step (bit 
flip) at a time, with all states sharing a given Hamming weight treated as 
equivalent, until it again reaches its starting state (see also 
Fig.~\ref{fig:Hamn=3}).

Given that the determination of the matrix permanent corresponds to an 
algebraic branching program from the state $|{\bf 0}\rangle$ to itself, 
effecting the spin transitions in the opposite direction (i.e.\ reversing the 
arrows in Fig.~\ref{fig:Hamn=3}) corresponds to taking the adjoint (complex
conjugate transpose) of $\tilde{M}$ or $\breve{M}$. Eq.~(\ref{eq:M0}) then
becomes 
\begin{equation}
\left(\tilde{M}^{\dag}\right)^{n+1}|{\bf 0}\rangle
=\left(\breve{M}^{\dag}\right)^n|{\bf 0}\rangle=P_n^*|{\bf 0}\rangle.
\end{equation}
One can then construct Hermitian operators
\begin{eqnarray}
\tilde{M}_R&=&\tilde{M}^{n+1}+\left(\tilde{M}^{\dag}\right)^{n+1},
\label{eq:Mherm1}\\
\tilde{M}_I&=&i\left[\tilde{M}^{n+1}-\left(\tilde{M}^{\dag}\right)^{n+1}\right],
\label{eq:Mherm2}
\end{eqnarray}
satisfying the eigenvalue equations
\begin{eqnarray}
\tilde{M}_R|{\bf 0}\rangle=\mbox{Re}\left(P_n\right)|{\bf 0}\rangle;\\
\tilde{M}_I|{\bf 0}\rangle=\mbox{Im}\left(P_n\right)|{\bf 0}\rangle.
\end{eqnarray}
Similar expressions apply to $\breve{M}$. While the 
operators~(\ref{eq:Mherm1}) and (\ref{eq:Mherm2}) are arguably more physical, 
their experimental realization could remain challenging given the complexity of 
the description, Eq.~(\ref{eq:Mn}). Also, unlike the case for $\tilde{M}^{n+1}$ 
or $\breve{M}^n$ alone, the non-zero eigenvalues for the 
remaining blocks of ~(\ref{eq:Mherm1}) and (\ref{eq:Mherm2}) are different from
$\mbox{Re}\left(P_n\right)$ and $\mbox{Im}\left(P_n\right)$.

\subsection{Classical Algorithm for the permanent}
\label{subsec:Palg}

While the result~(\ref{eq:M0}) is a statement about the eigenvalues, it 
suggests a straightforward approach to the calculation of the permanent without 
needing to determine the spectrum of $\tilde{M}$ or $\breve{M}$. Rather, one 
must only compute 
\begin{equation}
\tilde{M}^n|{\bf 0}\rangle=P_n|{\bf 1}\rangle\;\mbox{or}\;
\tilde{M}^{n+1}|{\bf 0}\rangle=P_n|{\bf 0}\rangle;\;\mbox{or}\;
\breve{M}^n|{\bf 0}\rangle=P_n|{\bf 0}\rangle.
\label{eq:easyperm}
\end{equation}
In other words, apply $\tilde{M}$ or $\breve{M}$ successively to the state 
$|{\bf 0}\rangle$ until all the amplitude is again concentrated on the state 
$|{\bf 0}\rangle$, and read out the result. 

The algorithm for the permanent then corresponds to an $n$-fold or $(n-1)$-fold
product of matrices with dimension $\left({n\atop i+1}\right)\times
\left({n\atop i}\right)$ ($i=\{0,1,\ldots,n-1\}$). Each column of the $i$th
matrix contains exactly $n-i$ non-zero elements, so that the matrices are
exponentially sparse. The total number of operations (multiplications and 
additions) is
\begin{equation}
\sum_{i=1}^n\left({n\atop i}\right)(2i)=n2^n.
\end{equation}
In comparison, Ryser's algorithm requires a total of $n2^{n+1}-(n+1)^2\sim 2n2^n$ operations 
for large $n$~\cite{Niu2020}. The scaling of the number of operations in the 
present case therefore matches that of the fastest-known algorithm, with a 
straightforward implementation, which could make it useful for practical applications.

\section{Fermionic and bosonic representations}
\label{sec:mapping}

The spin model~(\ref{eq:Ham}) can be naturally represented in terms of 
Schwinger bosons, and fermions via the Jordan-Wigner transformation. These are 
discussed in the next two subsections.

\subsection{Bosons}
\label{subsec:bosons}

Spin-$1/2$ particles can be mapped to Schwinger bosons as follows:
\begin{eqnarray}
\sigma_j^+&=&\sigma_j^x-i\sigma_j^y=a_j^{\dag}b_j;\nonumber \\
\sigma_j^-&=&\sigma_j^x+i\sigma_j^y=b_j^{\dag}a_j;\nonumber \\
\sigma_j^z&=&a^{\dag}_ja_j-b^{\dag}_jb_j.
\end{eqnarray}
Each spin operator therefore involves two `species' of bosons, satisfying the
relations
\begin{equation}
[a_i,a_j^{\dag}]=\delta_{ij};\quad[a_i,a_j] =[a_i^{\dag},a_j^{\dag}]=0
\end{equation}
and likewise for $b$-species bosons,
where $[x,y]=xy-yx$ is the commutator. These are supplemented with the 
unit-occupancy condition
\begin{equation}
a_j^{\dag}a_j+b^{\dag}_jb_j=1,
\end{equation}
which specifies that each site is occupied by exactly one boson of either 
species. The Schwinger approach therefore maps spins to hard-core two-species 
bosons at exactly unit filling. The model~(\ref{eq:Ham}) expressed in terms of 
Schwinger bosons is then
\begin{equation}
\tilde{M}_b=\sum_{\bf i}\sum_{j=0}^{n-1}w_{h({\bf i}),j}a_j^{\dag}b_j
|{\bf i}\rangle\langle{\bf i}|+\prod_jb_j^{\dag}a_j,
\label{eq:Hambosons}
\end{equation}
where the bit in the string ${\bf i}$ is unity (zero) if occupied by a boson
of species $a$ ($b)$, and the zero state is
\begin{equation}
|{\bf 0}\rangle=\prod_{j=0}^{n-1}b_j^{\dag}|{\mathcal O}\rangle.
\end{equation}
The graph associated with $\tilde{M}_b$ is indistinguishable from that of
$\tilde{M}$, i.e.\ Fig.~\ref{fig:Hamn=3} for $n=3$.

It is instructive to write the action of the $n$th power of $\tilde{M}_b$ on 
the zero state:
\begin{eqnarray}
\tilde{M}_b^n|{\bf 0}\rangle&=&
\left(\sum_{j=0}^{n-1}w_{n-1,j}a_j^{\dag}b_j\right)
\left(\sum_{j=0}^{n-1}w_{n-2,j}a_j^{\dag}b_j\right)\nonumber \\
&\times&\cdots\times\left(\sum_{j=0}^{n-1}w_{0,j}a_j^{\dag}b_j\right)
\prod_{j=0}^{n-1}b_j^{\dag}|{\mathcal O}\rangle \nonumber \\
&=&\left(\sum_{j=0}^{n-1}w_{n-1,j}a_j^{\dag}\right)
\left(\sum_{j=0}^{n-1}w_{n-2,j}a_j^{\dag}\right)\nonumber \\
&\times&\cdots\times\left(\sum_{j=0}^{n-1}w_{0,j}a_j^{\dag}\right)
|{\mathcal O}\rangle.
\label{eq:Hamnbosons}
\end{eqnarray}
In the second line above, all operators for the $b$-species bosons can be 
omitted because each creation of an $a$-species boson must be accompanied by
the annihilation of a $b$-species boson, and after $n$ powers of $\tilde{M}_b$
all sites have been accounted for. Furthermore, the hard-core condition acts in
the same way as a Pauli exclusion principle: if a $b$-species boson occupies 
site $j$, then the $b_j^{\dag}$ operator returns zero. Expansion of the terms 
in Eq.~(\ref{eq:Hamnbosons}) then returns the permanent because the $b$-species 
bosons all commute. 

\subsection{Fermions}
\label{subsec:fermions}

The Jordan-Wigner transformation corresponds to mapping the spin operators to 
`spinless' fermions:
\begin{eqnarray}
\sigma_j^+&=&\exp\left(i\pi\sum_{k=j+1}^{n-1}f_k^{\dag}f_k\right)f_j^{\dag};
\nonumber \\
\sigma_j^-&=&\exp\left(i\pi\sum_{k=j+1}^{n-1}f_k^{\dag}f_k\right)f_j;
\nonumber \\
\sigma_j^z&=&2f_j^{\dag}f_j-1,
\label{eq:JW}
\end{eqnarray}
where the site-dependent fermionic creation ($f^{\dag}_j$) and annihilation
($f_j$) operators satisfy the anticommutation relations
\begin{equation}
\{f_i,f_j^{\dag}\}=\delta_{ij};\quad\{f_i,f_j\} =\{f_i^{\dag},f_j^{\dag}\}=0,
\end{equation}
and $\{x,y\}=xy+yx$. The first of these automatically ensures the Pauli
condition forbidding double occupancy of sites; thus, basis states can 
therefore again be indexed by bitstrings ${\bf i}$, but now where $0$ ($1$)
signifies the absence (presence) of a fermion at position $j$. Canonical 
ordering is assumed, where creation operators appear with indices in descending 
order; for example 
\begin{equation}
|1010\rangle=f_2^{\dag}f_0^{\dag}|{\mathcal O}\rangle,
\end{equation}
where $|{\mathcal O}\rangle$ denotes the particle vacuum.

\begin{figure}[t]
\begin{tikzpicture}[decoration={
    markings,
    mark=at position 0.35 with {\arrow{>}}},
		scale=0.75,every node/.style={scale=0.75}
]


\draw[fill] (2.55, 2.0) arc(0:360:0.05) -- cycle;
\draw[fill] (0.55, 0.0) arc(0:360:0.05) -- cycle;
\draw[fill] (2.55, 0.0) arc(0:360:0.05) -- cycle;
\draw[fill] (4.55, 0.0) arc(0:360:0.05) -- cycle;
\draw[fill] (0.55, -2.0) arc(0:360:0.05) -- cycle;
\draw[fill] (2.55, -2.0) arc(0:360:0.05) -- cycle;
\draw[fill] (4.55, -2.0) arc(0:360:0.05) -- cycle;
\draw[fill] (2.55, -4.0) arc(0:360:0.05) -- cycle;

\draw[thick,black,postaction={decorate}] (2.5,2) -- (0.5,0)
  node [midway,left] {\small $w_{0,0}$};
\draw[thick,black,postaction={decorate}] (2.5,2) -- (2.5,0)
  node [pos=0.7,right] {\small $w_{0,1}$};
\draw[thick,black,postaction={decorate}] (2.5,2) -- (4.5,0)
  node [midway,right] {\small $w_{0,2}$};
\draw[thick,black,postaction={decorate}] (0.5,0) -- (0.5,-2)
  node [midway,left] {\small $w_{1,1}$};
	\draw[thick,black,postaction={decorate}] (0.5,0) -- (2.5,-2)
  node [pos=0.3,above] {\small $w_{1,2}$};
\draw[thick,black,postaction={decorate}] (2.5,0) -- (0.5,-2)
  node [near start,above] {\small $-w_{1,0}$};
\draw[thick,black,postaction={decorate}] (2.5,0) -- (4.5,-2)
  node [pos=0.3,above] {\small $w_{1,2}$};
\draw[thick,black,postaction={decorate}] (4.5,0) -- (4.5,-2)
  node [midway,right] {\small $-w_{1,1}$};
\draw[thick,black,postaction={decorate}] (4.5,0) -- (2.5,-2)
  node [near start,above] {\small $-w_{1,0}$};
\draw[thick,black,postaction={decorate}] (0.5,-2) -- (2.5,-4)
  node [midway,left] {\small $w_{2,2}$};
\draw[thick,black,postaction={decorate}] (2.5,-2) -- (2.5,-4)
  node [pos=0.2,right] {\small $-w_{2,1}$};
\draw[thick,black,postaction={decorate}] (4.5,-2) -- (2.5,-4)
  node [midway,right] {\small $w_{2,0}$};

\draw[thick,black] (2.5,2.2) node {$|000\rangle$};
\draw[thick,black] (0.05,0.0) node {$|100\rangle$};
\draw[thick,black] (2.9,-0.0) node {$|010\rangle$};
\draw[thick,black] (4.95,0.0) node {$|001\rangle$};
\draw[thick,black] (0.05,-2.0) node {$|110\rangle$};
\draw[thick,black] (2.9,-2.0) node {$|101\rangle$};
\draw[thick,black] (4.95,-2.0) node {$|011\rangle$};
\draw[thick,black] (2.5,-4.2) node {$|000\rangle$};


\draw[] (2.5,-4.8) node {(a)};


\draw[fill] (8.55, 2.0) arc(0:360:0.05) -- cycle;
\draw[fill] (6.55, 0.0) arc(0:360:0.05) -- cycle;
\draw[fill] (8.55, 0.0) arc(0:360:0.05) -- cycle;
\draw[fill] (10.55, 0.0) arc(0:360:0.05) -- cycle;
\draw[fill] (6.55, -2.0) arc(0:360:0.05) -- cycle;
\draw[fill] (8.55, -2.0) arc(0:360:0.05) -- cycle;
\draw[fill] (10.55, -2.0) arc(0:360:0.05) -- cycle;
\draw[fill] (8.55, -4.0) arc(0:360:0.05) -- cycle;

\draw[thick,black,postaction={decorate}] (8.5,2) -- (6.5,0)
  node [midway,left] {$w_{0,0}'$};
\draw[thick,black,postaction={decorate}] (8.5,2) -- (8.5,0)
  node [pos=0.7,right] {$w_{0,1}'$};
\draw[dashed,black,postaction={decorate}] (8.5,2) -- (10.5,0)
  node [midway,right] {\small $w_{0,2}$};
\draw[thick,black,postaction={decorate}] (6.5,0) -- (6.5,-2)
  node [midway,left] {$w_{1,1}'$};
\draw[dashed,black,postaction={decorate}] (6.5,0) -- (8.5,-2)
  node [pos=0.3,above] {\small $w_{1,2}$};
\draw[thick,black,postaction={decorate}] (8.5,0) -- (6.5,-2)
  node [near start,above] {$-w_{1,0}'$};
\draw[dashed,black,postaction={decorate}] (8.5,0) -- (10.5,-2)
  node [pos=0.3,above] {\small $w_{1,2}$};
\draw[dashed,black,postaction={decorate}] (10.5,0) -- (10.5,-2)
  node [pos=0.3,right] {\small $-w_{1,1}$};
\draw[dashed,black,postaction={decorate}] (10.5,0) -- (8.5,-2)
  node [near start,above] {\small $-w_{1,0}$};
\draw[dashed,black,postaction={decorate}] (10.5,-2) -- (8.5,-4)
  node [pos=0.3,right] {\small $w_{2,0}$};
\draw[dashed,black,postaction={decorate}] (8.5,-2) -- (8.5,-4)
  node [pos=0.3,right] {\small $-w_{2,1}$};
\draw[thick,black,postaction={decorate}] (6.5,-2) -- (8.5,-4)
  node [midway,left] {$w_{2,2}$};

\draw[thick,black] (8.5,2.2) node {$|000\rangle$};
\draw[thick,black] (6.05,0.0) node {$|100\rangle$};
\draw[thick,black] (8.9,-0.0) node {$|010\rangle$};
\draw[thick,black] (10.95,0.0) node {$|001\rangle$};
\draw[thick,black] (6.05,-2.0) node {$|110\rangle$};
\draw[thick,black] (8.9,-2.0) node {$|101\rangle$};
\draw[thick,black] (10.95,-2.0) node {$|011\rangle$};
\draw[thick,black] (8.5,-4.2) node {$|000\rangle$};


\draw[] (8.5,-4.8) node {(b)};


\draw[fill] (2.55, -6.0) arc(0:360:0.05) -- cycle;
\draw[fill] (0.55, -8.0) arc(0:360:0.05) -- cycle;
\draw[fill] (0.55, -10.0) arc(0:360:0.05) -- cycle;
\draw[fill] (2.55, -8.0) arc(0:360:0.05) -- cycle;
\draw[fill] (2.55, -12.0) arc(0:360:0.05) -- cycle;

\draw[thick,black,postaction={decorate}] (2.5,-6) -- (0.5,-8)
  node [midway,left] {$w_{0,0}'$};
\draw[thick,black,postaction={decorate}] (2.5,-6) -- (2.5,-8)
  node [pos=0.7,right] {$w_{0,1}'$};
\draw[thick,black,postaction={decorate}] (2.5,-8) -- (0.5,-10)
  node [near start,above] {$-w_{1,0}'$};
\draw[thick,black,postaction={decorate}] (0.5,-8) -- (0.5,-10)
  node [midway,left] {$w_{1,1}'$};
\draw[thick,black,postaction={decorate}] (0.5,-10) -- (2.5,-12)
  node [midway,left] {$w_{2,2}'$};

\draw[thick,black] (2.5,-5.8) node {$|000\rangle$};
\draw[thick,black] (0.05,-8.0) node {$|100\rangle$};
\draw[thick,black] (0.05,-10.0) node {$|110\rangle$};
\draw[thick,black] (2.9,-8.0) node {$|010\rangle$};
\draw[thick,black] (2.5,-12.2) node {$|000\rangle$};


\draw[thick,black] (2.5,-12.8) node {(c)};


\draw[fill] (8.55, -6.0) arc(0:360:0.05) -- cycle;
\draw[fill] (6.55, -8.0) arc(0:360:0.05) -- cycle;
\draw[fill] (6.55, -10.0) arc(0:360:0.05) -- cycle;
\draw[fill] (8.55, -8.0) arc(0:360:0.05) -- cycle;
\draw[fill] (8.55, -12.0) arc(0:360:0.05) -- cycle;

\draw[thick,black,postaction={decorate}] (8.5,-6) -- (6.5,-8)
  node [midway,left] {$w_{0,0}''$};
\draw[dashed,black,postaction={decorate}] (8.5,-6) -- (8.5,-8)
  node [pos=0.7,right] {\small $w_{0,1}'$};
\draw[dashed,postaction={decorate}] (8.5,-8) -- (6.5,-10)
  node [near start,above] {\small $-w_{1,0}'$};
\draw[thick,black,postaction={decorate}] (6.5,-8) -- (6.5,-10)
  node [midway,left] {$w_{1,1}'$};
\draw[thick,black,postaction={decorate}] (6.5,-10) -- (8.5,-12)
  node [midway,left] {$w_{2,2}$};

\draw[thick,black] (8.5,-5.8) node {$|000\rangle$};
\draw[thick,black] (6.05,-8.0) node {$|100\rangle$};
\draw[thick,black] (6.05,-10.0) node {$|110\rangle$};
\draw[thick,black] (8.9,-8.0) node {$|010\rangle$};
\draw[thick,black] (8.5,-12.2) node {$|000\rangle$};


\draw[] (8.5,-12.8) node {(d)};

\end{tikzpicture}
\caption{Depictions of the fermionic model $\breve{M}_{f,{\rm alt}}$, 
Eq.~(\ref{eq:Hamfermionsalt}) with $|111\rangle\to|000\rangle$, for $n=3$. Its 
original form is shown in (a), while various stages of row reduction are shown 
in (b)-(d).}
\label{fig:Hamn=3fermions}
\end{figure}
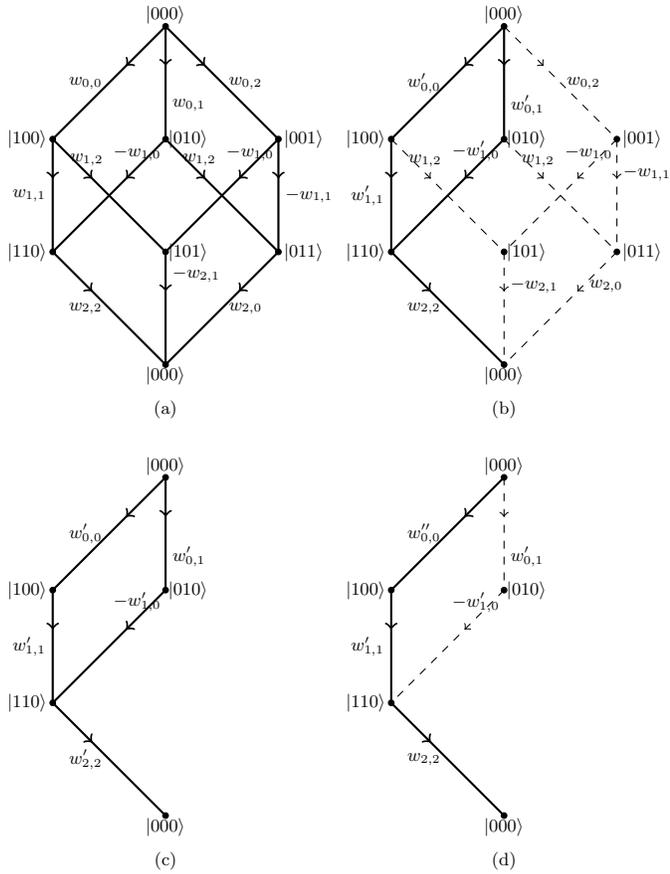

The phases appearing in Eq.~(\ref{eq:JW}) ensure that the fermions anticommute
on all sites; alternatively, they ensure the normal / canonical ordering of 
basis states. For example:
\begin{eqnarray}
f_0^{\dag}|1010\rangle&=&f_0^{\dag}f_2^{\dag}f_0^{\dag}|{\mathcal O}
\rangle=0;\nonumber \\
f_1^{\dag}|1010\rangle&=&f_1^{\dag}f_2^{\dag}f_0^{\dag}|{\mathcal O}
\rangle=-f_2^{\dag}f_1^{\dag}f_0^{\dag}|{\mathcal O}\rangle=-|1110\rangle;
\nonumber \\
f_3^{\dag}|1010\rangle&=&f_3^{\dag}f_2^{\dag}f_0^{\dag}|{\mathcal O}
\rangle=|1011\rangle.
\end{eqnarray}
The Jordan-Wigner transformation~(\ref{eq:JW}) counts the number of fermions to 
the right of (i.e.\ with index greater than) where the spin is flipped / 
fermion is created, and multiplies the transition amplitude by $-1$ if this 
number is odd. In this way, the negative signs arising from the fermionic 
anticommutation are cancelled and the transition amplitudes all remain 
positive. The model~(\ref{eq:Ham}) expressed in terms of fermions then becomes
\begin{equation}
\tilde{M}_{\rm JW}=\sum_{\bf i}\sum_{j=0}^{n-1}w_{h({\bf i}),j}s_{{\bf i},j}
f_j^{\dag}|{\bf i}\rangle\langle{\bf i}|+\prod_jf_j,
\label{eq:HamJW}
\end{equation}
where the function $s_{{\bf i},j}$ incorporates the Jordan-Wigner phases for
creation of a fermion at position $j$ on a basis state with occupation indexed
by occupation state $|{\bf i}\rangle$ defined by bitstring ${\bf i}$. 

An explicit example is shown for the $n=3$ case in 
Fig.~\ref{fig:Hamn=3fermions}(a). Consider the $|100\rangle\to|110\rangle$ and
$|010\rangle\to|110\rangle$ transitions. For the former transition, a fermion
is created in site 1, to the right of a fermion already in site 0, so there is 
no additional Jordan-Wigner phase; likewise, the final state 
$|110\rangle=f_1^{\dag}f_0^{\dag}|{\mathcal O}\rangle$ is already normal 
ordered. Thus, the edge weight $w_{1,1}$ remains unchanged. For the latter 
transition, the fermion created in site 0 is to the left of a fermion already 
in site 1, which yields a negative contribution from the Jordan-Wigner 
transformation, reflected in the signed edge weight $-w_{1,0}$ in 
Fig.~\ref{fig:Hamn=3fermions}(a). At the same time, the final state 
$|110\rangle=f_0^{\dag}f_1^{\dag}|{\mathcal O}\rangle$ requires one fermionic 
anticommutation to bring it back to normal ordering, which cancels the negative
sign and effectively restores the total edge weight to its original value in
the spin representation. Thus, within the context of a binary branching
process, the sum of the path weights of Fig.~\ref{fig:Hamn=3fermions}(a) still
constitute the permanent, despite the appearance of signed edge weights.

To construct an algebraic branching program for true fermions one either must
maintain all edge weights and keep track of the fermionic anticommutation 
relations defining the occupation states, in which case the model is
\begin{equation}
\tilde{M}_f=\sum_{\bf i}\sum_{j=0}^{n-1}w_{h({\bf i}),j}f_j^{\dag}
|{\bf i}\rangle\langle{\bf i}|+\prod_jf_j
\label{eq:Hamfermions}
\end{equation}
and $|{\bf i}\rangle$ represent occupation states; or one must account for all 
Jordan-Wigner phases to appropriately sign all edge weights but treat the 
states $|{\bf i}\rangle$ instead as ordinary bitstrings, in which case the 
model is instead
\begin{equation}
\tilde{M}_{f,{\rm alt}}=\sum_{\bf i}\sum_{j=0}^{n-1}w_{h({\bf i}),j}
s_{{\bf i},j}\sigma^+_j|{\bf i}\rangle\langle{\bf i}|+\prod_j\sigma^-_j,
\label{eq:Hamfermionsalt}
\end{equation}
and now the $\sigma^{\pm}$ are interpreted as classical bit-flip operators.
When expressing the fermionic model in terms of creation and annihilation
operators, Eq.~(\ref{eq:Hamfermions}) is preferable, but 
Eq.~(\ref{eq:Hamfermionsalt}) is more convenient in the graph adjacency matrix 
representation. Now, Fig.~\ref{fig:Hamn=3fermions}(a) depicts a truly signed 
binary branching
process, and the sum of the path weights constitute the determinant, rather 
than the permanent, of $M$ for $n=3$. The $w_{0,0}w_{1,1}w_{2,2}$ path serves 
as the reference, where the second indices for the weights in this product 
constitute the integer list $\{012\}$. All other paths are characterized by an 
overall minus (plus) sign if the integer list derived from the second index of 
the weights for that path corresponds to an even (odd) number of inversions 
of the reference list; for example, the odd permutations $\{021\}$, $\{102\}$, 
and $\{210\}$ correspond to paths with negative total weights 
$-w_{0,0}w_{1,2}w_{2,1}$, $-w_{0,1}w_{1,0}w_{2,2}$, and 
$-w_{0,2}w_{1,1}w_{2,0}$, respectively. Thus, one expects that the eigenvalues 
of the operator~(\ref{eq:Hamfermions}) are the determinant, as will be 
discussed further below.

Another noteworthy property of the fermionic graph is that it is unbalanced: 
there is no vertex sign switching that can remove all of the minus 
signs~\cite{Zaslavsky1982}; alternatively
expressed, there is no diagonal matrix whose entries are $\{1,-1\}$ that can 
map Eq.~(\ref{eq:Hamfermions}) to a form without any $s_{{\bf i},j}$ factors. 
(This is another way of stating that the determinant derived in this way cannot 
be mapped to the permanent by a local unitary, though this is already 
obvious as unitary transformations preserve the eigenvalues). There remains
the intriguing possibility that there is a non-unitary operator that can effect
the map, but this is not explored in the present work. 

Similarly, it is not possible to map existing weights to their negatives in
order to map the determinant to the permanent. For the $n=3$ case, one could
reassign $w_{1,2}\to -\bar{w}_{1,2}$ and $w_{2,1}\to -\bar{w}_{2,1}$ to 
remove the negative signs on all edges with these labels in 
Fig.~\ref{fig:Hamn=3fermions}(a), but this still leaves signs on edges labeled
by $w_{1,1}$ which cannot be removed.

As in the bosonic case, it is worthwhile to express the action of the $n$th 
power of $\tilde{M}_{f,{\rm alt}}$ on the vacuum state:
\begin{eqnarray}
\tilde{M}_f^n|{\bf 0}\rangle&=&
\left(\sum_{j=0}^{n-1}w_{n-1,j}f_j^{\dag}\right)
\left(\sum_{j=0}^{n-1}w_{n-2,j}f_j^{\dag}\right)\nonumber \\
&\times&\cdots\times\left(\sum_{j=0}^{n-1}w_{0,j}f_j^{\dag}\right)
|{\mathcal O}\rangle.
\label{eq:Hamnfermions}
\end{eqnarray}
This simple and apparently separable representation for the output state, as 
products of similar terms, is possible because of the Pauli principle and the 
anticommutation relations: any attempted creation of a fermion in an 
already-occupied site is zero, and the signs of the final many-fermion states
will reflect the number of permutations required to express them in normal
ordering. Furthermore, the result is clearly the determinant $D_n$ (or its 
negative) rather than the permanent. The states~(\ref{eq:Hamnfermions}) and 
(\ref{eq:Hamnbosons}) reveal a close connection between the determinant and 
the permanent expressed in terms of indistinguishable particles. 

Consider explicitly the $n=3$ case:
\begin{eqnarray}
\tilde{M}_f^3|{\bf 0}\rangle&=&\left(w_{2,0}f_0^{\dag}+w_{2,1}f_1^{\dag}
+w_{2,2}f_2^{\dag}\right)\nonumber\\
&\times&\left(w_{1,0}f_0^{\dag}+w_{1,1}f_1^{\dag}+w_{1,2}f_{2}^{\dag}\right)
\nonumber\\
&\times&\left(w_{0,0}f_0^{\dag}+w_{0,1}f_1^{\dag}+w_{0,2}f_{2}^{\dag}\right)
|{\mathcal O}\rangle.\nonumber \\
&=&w_{2,2}f_2^{\dag}\left(w_{1,1}f_1^{\dag}w_{0,0}f_0^{\dag}+w_{1,0}f_0^{\dag}
w_{0,1}f_1^{\dag}\right)\nonumber \\
&+&w_{2,1}f_1^{\dag}\left(w_{1,2}f_{2}^{\dag}w_{0,0}f_0^{\dag}
+w_{1,0}f_0^{\dag}w_{0,2}f_{2}^{\dag}\right)\nonumber \\
&+&w_{2,0}f_0^{\dag}\left(w_{1,1}f_1^{\dag}w_{0,2}f_{2}^{\dag}
+w_{1,2}f_{2}^{\dag}w_{0,1}f_1^{\dag}\right)|{\mathcal O}\rangle\nonumber \\
&=&\left[w_{2,2}\left(w_{1,1}w_{0,0}-w_{1,0}w_{0,1}\right)\right.\nonumber \\
&+&w_{2,1}\left(-w_{1,2}w_{0,0}+w_{1,0}w_{0,2}\right)\nonumber \\
&+&\left.w_{2,0}\left(-w_{1,1}w_{0,2}+w_{1,2}w_{0,1}\right)\right]
f_2^{\dag}f_1^{\dag}f_0^{\dag}|{\mathcal O}\rangle\nonumber \\
&=&D_3f_2^{\dag}f_1^{\dag}f_0^{\dag}|{\mathcal O}\rangle.
\label{eq:Hamfermionn=3}
\end{eqnarray}
Recapitulating the arguments of Sec.~\ref{subsec:eigensystem} but for 
$\tilde{M}_f$ instead of $\tilde{M}$ or 
$\breve{M}$, one obtains that the only non-zero eigenvalues of $\tilde{M}_f$ 
are given by the $(n+1)$th roots of $D_n$.

\section{Row reductions}
\label{sec:rowreduce}

The exponentially small rank of the matrices $\tilde{M}$ and $\breve{M}$,
discussed in Section~\ref{subsec:eigensystem}, suggests that it might be 
possible to apply row reductions to reduce their dimension without
affecting the non-zero eigenvalues. Just as Gaussian elimination reduces a 
matrix to upper (or lower) triangular form, so that the determinant (which 
would otherwise require summing $n!$ terms) can be evaluated by a product of 
the diagonal elements, row reduction of $\tilde{M}$ or $\breve{M}$ reduces the 
$n!$ paths of the algebraic branching program to a single path by deleting 
vertices and reweighting edges. As shown below, row reductions in the fermionic 
model correspond precisely to the Gaussian elimination approach to 
evaluating the determinant. The bosonic version provides a roadmap for row 
reductions to evaluate the permanent, but doesn't appear to provide a speedup
over the direct matrix multiplication method discussed in 
Sec.~\ref{subsec:Palg}.

As shown in Sec.~\ref{subsec:eigensystem}, the blocks $\tilde{M}_m$ of 
$\tilde{M}^{n+1}$ and $\breve{M}^n$ have dimension 
$\left({n\atop m}\right)$ but are all unit rank, so that the ranks of 
$\tilde{M}^{n+1}$ and $\breve{M}^n$ are $n+1$ and $n$, respectively. In 
contrast, the matrices $\tilde{M}$ and $\breve{M}$ are not block diagonal, and 
their eigenvectors are no longer given by $\tilde{M}^m|{\bf 0}\rangle$ and 
$\breve{M}^m|{\bf 0}\rangle$, respectively. Consider $\breve{M}$ for
concreteness. While the $n$ non-zero eigenvalues correspond to the $n$th roots 
of the permanent, the zero eigenvalues have multiplicity $2^{n-1}-n$; thus, the
kernel of $\breve{M}$ is comprised of generalized zero eigenvectors of rank 1 
up to $n-1$. The set of linearly independent vectors spanning these defective 
matrices must therefore be obtained sequentially. The standard procedure is to 
obtain the set of $r_m$ generalized zero rank-$m$ vectors $|v_i^{(m)}\rangle$, 
$1\leq i\leq r_m$, such that $\breve{M}^m|v_i^{(m)}\rangle
=|{\mathcal O}\rangle$. The rank-nullity theorem ensures that 
$n+\sum_{m=1}^{n-1}r_m=2^n-1$.

In practice there is a more efficient iterative procedure to obtain the kernel. 
First generate the reduced row echelon form (also known as the pivot matrix) 
$B_1$ for $\breve{M}$, via Gauss-Jordan elimination. For any rank-deficient 
matrix such as $\breve{M}$, the deviation of $B_1$ from the identity is driven 
entirely by the $r_1$ rank-1 zero eigenvectors; thus the 
$(2^n-1-r_1)\times (2^n-1)$ matrix $B_1$ annihilates the null space: 
$B_1|v_i^{(1)}\rangle=0$. One can then find an 
$n\times(2^n-1-r_1)$ matrix $A_1$ such that $\breve{M}=A_1B_1$; its matrix 
elements coincide with those of $\breve{M}$ but with $r_1$ columns removed 
whose indices correspond to the location of the first non-zero element of each 
$|v_i^{(1)}\rangle$. The $(2^n-1-r_1)\times(2^n-1-r_1)$ matrix 
$\breve{M}^{(2)}\equiv B_1A_1$ therefore has the same eigenvalues as 
$\breve{M}$ but now with $r_1$ fewer zeroes.

The rank-2 generalized eigenvectors are the solutions of 
\begin{equation}
\breve{M}^2|v_i^{(2)}\rangle=A_1B_1A_1B_1|v_i^{(2)}\rangle
=|0\rangle,
\label{eq:rank2a}
\end{equation}
for $1\leq i\leq r_2$, which can be rewritten as
\begin{equation}
A_1\breve{M}^{(2)}\left(B_1|v_i^{(2)}\rangle\right)=0.
\label{eq:rank2b}
\end{equation}
At the same time, the zero eigenvectors of $\breve{M}^{(2)}$ are the solutions
of 
\begin{equation}
\breve{M}^{(2)}|\tilde{v}_i^{(2)}\rangle=0.
\label{eq:rank2c}
\end{equation}
Thus, with the identification 
$|\tilde{v}_i^{(2)}\rangle\equiv B_1|v_i^{(2)}\rangle$,
Eq.~(\ref{eq:rank2b}) is automatically satisfied by Eq.~(\ref{eq:rank2c}),
and solving the latter is more efficient than the former due to the smaller 
matrix dimension. It is straightforward to verify that the non-zero eigenstates 
of interest from Eq.~(\ref{eq:eigenvectors}),
\begin{equation}
|\phi_n^{(1)}(k)\rangle=e^{-i2\pi k/n}\sum_{j=0}^{n-1}e^{i2\pi jk/n}
\left(\frac{\breve{M}}{P_n^{1/n}}\right)^j|0^{\otimes n}\rangle,
\end{equation}
are transformed into
\begin{eqnarray}
|\phi_n^{(2)}(k)\rangle&=&e^{-i2\pi k/n}\sum_{j=0}^{n-1}e^{i2\pi jk/n}
\left(\frac{\breve{M}^{(2)}}{P_n^{1/n}}\right)^j|0^{\otimes n}\rangle
\nonumber \\
&=&B_1|\phi_n^{(1)}(k)\rangle.
\end{eqnarray}

The procedure is then repeated for $\breve{M}^{(2)}=A_2B_2$. After $n-1$ 
iterations, the original rank-deficient $(2^n-1)$-dimensional matrix 
$\breve{M}$ is reduced to a full-rank $n$-dimensional matrix 
$\breve{M}^{(n-1)}$ with eigenvectors
$\prod_{i=1}^{n-1}B_{n-i}|\phi_n(k)\rangle$ and corresponding eigenvalues 
$P_n^{1/n}$. As shown below, this procedure is equivalent to Gaussian 
elimination of $M$ for the evaluation of the determinant, and also provides an 
equivalent systematic approach to the evaluation of the permanent.

\subsection{Example: Three Fermions}
\label{subsec:rowreducefermions}

Consider row reductions of $\tilde{M}_{f,{\rm alt}}$, 
Eq.~(\ref{eq:Hamfermionsalt}), for the specific case $n=3$, depicted in 
Fig.~\ref{fig:Hamn=3fermions}). The (unnormalized) rank-1 generalized zero 
eigenvectors can be written as
\begin{eqnarray}
|v_1^{(1)}\rangle&=&|001\rangle+\frac{w_{1,1}}{w_{1,2}}|010\rangle
+\frac{w_{1,0}}{w_{1,2}}|100\rangle;\nonumber \\
|v_2^{(1)}\rangle&=&|011\rangle-\frac{w_{2,0}}{w_{2,2}}|110\rangle;\nonumber \\
|v_3^{(1)}\rangle&=&|101\rangle+\frac{w_{2,1}}{w_{2,2}}|110\rangle,
\label{eq:v1f}
\end{eqnarray}
so that one may eliminate vertices labeled by the bitstrings $001$, $011$, and 
$101$. The matrix $B_1$ must satisfy $B_1|v_i^{(1)}\rangle
=0$; a sufficient construction is 
\begin{eqnarray}
B_1&=&I-|v_1^{(1)}\rangle\langle 001|-|v_2^{(1)}\rangle\langle 011|
-|v_3^{(1)}\rangle\langle 101|\nonumber \\
&=&\begin{pmatrix}
1 & 0 & 0 & 0 & 0 & 0 & 0\cr
0 & -\frac{w_{1,1}}{w_{1,2}} & 1 & 0 & 0 & 0 & 0\cr
0 & -\frac{w_{1,0}}{w_{1,2}} & 0 & 0 & 1 & 0 & 0\cr
0 & 0 & 0 & \frac{w_{2,0}}{w_{2,2}} & 0 & -\frac{w_{2,1}}{w_{2,2}} & 1\cr
\end{pmatrix},
\label{eq:B1}
\end{eqnarray}
where rows and columns indices are labeled by bitstrings with the least
significant bit on the right. Here, $B_1$ is expressed in the somewhat
unconventional lower-triangular reduced row echelon form. Likewise,
\begin{eqnarray}
A_1&=&\breve{M}_{f,{\rm alt}}\backslash \{\langle 001|, \langle 011|,
\langle 101|\}\nonumber \\
&=&\begin{pmatrix}
0 & 0 & 0 & w_{2,2}\cr
w_{0,2} & 0 & 0 & 0\cr
w_{0,1} & 0 & 0 & 0\cr
0 & w_{1,2} & 0 & 0\cr
w_{0,0} & 0 & 0 & 0\cr
0 & 0 & w_{1,2} & 0\cr
0 & -w_{1,0} & w_{1,1} & 0\cr
\end{pmatrix}.
\end{eqnarray}
It is straightforward to verify that $A_1B_1=\breve{M}_{f,{\rm alt}}$. One then 
obtains
\begin{equation}
\tilde{M}_{f,{\rm alt}}^{(2)}=B_1A_1=\begin{pmatrix}
0 & 0 & 0 & w_{2,2}\cr
w_{0,1}' & 0 & 0 & 0\cr
w_{0,0}' & 0 & 0 & 0\cr
0 & -w_{1,0}' & w_{1,1}' & 0\cr
\end{pmatrix},
\end{equation}
where $w_{0,0}'$, $w_{0,1}'$, $w_{1,0}'$, and $w_{1,1}'$ coincide with the 
reweighted terms in $M$ derived from a first round of Gaussian elimination,
defined in Eqs.~(\ref{eq:primed}) and (\ref{eq:primed2}). 

It is illuminating to view this first round as an operation on the graph
representing the binary branching process, as depicted in 
Figs.~\ref{fig:Hamn=3fermions}(b) and (c). Vertices labeled by bitstrings 
$001$, $011$, and $101$ are deleted, reducing the total number of branches from
six to two. The contributions to the determinant of the branches through the
deleted vertices are incorporated by reweighting remaining edges. For example, 
the weight $-w_{1,2}w_{2,1}$ of the path from $|100\rangle$ to $|000\rangle$
through vertex $|101\rangle$ is incorporated into the new path weight 
$w_{1,1}'w_{2,2}$; similarly, the path from $|010\rangle$ to $|000\rangle$ 
through deleted vertex $|011\rangle$ is incorporated in $w_{1,0}'$. 
Perhaps surprisingly, these edge reweightings can also compensate for both of 
the deleted paths from $|001\rangle$ to $|000\rangle$ through the two deleted
vertices $|011\rangle$ and $|101\rangle$. Crucially, for fermions, the total 
path weights after the transformation are products of revised edge weights; as 
discussed in Sec.~\ref{subsec:permderm}, cancellation of signed terms ensure 
that the total weights for the reduced branching process still coincide with 
the determinant. 

The remaining (unnormalized) rank-2 generalized zero eigenvector can now be 
efficiently written as
\begin{equation}
|v^{(2)}\rangle=|010\rangle+\frac{w_{1,0}'}{w_{1,1}'}|100\rangle,
\end{equation}
so that one may eliminate the vertex labeled by the bitstring $010$. The matrix 
$B_2$ must satisfy $B_2|v^{(2)}\rangle=|{\mathcal O}\rangle$:
\begin{eqnarray}
B_2&=&I-|v^{(2)}\rangle\langle 010|\nonumber \\
&=&\begin{pmatrix}
1 & 0 & 0 & 0\cr
0 & -\frac{w_{1,0}'}{w_{1,1}'} & 1 & 0\cr
0 & 0 & 0 & 1\cr
\end{pmatrix}.
\end{eqnarray}
Likewise,
\begin{eqnarray}
A_2&=&\breve{M}_{f,{\rm alt}}^{(2)}\backslash\langle 010|\nonumber \\
&=&\begin{pmatrix}
0 & 0 & w_{2,2}\cr
w_{0,1}' & 0 & 0\cr
w_{0,0}' & 0 & 0\cr
0 & w_{1,1}' & 0\cr
\end{pmatrix}.
\end{eqnarray}
Again, it is straightforward to verify that $A_2B_2
=\breve{M}_{f,{\rm alt}}^{(2)}$. One 
then obtains
\begin{equation}
\breve{M}_{f,{\rm alt}}^{(3)}=B_2A_2=\begin{pmatrix}
0 & 0 & w_{2,2}\cr
w_{0,0}'' & 0 & 0\cr
0 & w_{1,1}' & 0\cr
\end{pmatrix},
\end{equation}
where $w_{0,0}''$ coincides with Eq.~(\ref{eq:w00pp}). The eigenvalues of 
$\breve{M}^{(3)}$ are the cube roots of $D_3=w_{0,0}''w_{1,1}'w_{2,2}$.
In this example, the second and final round of Gauss-Jordan elimination 
corresponds to deleting the vertex labeled by bitstring $010$, as depicted in
Fig.~\ref{fig:Hamn=3fermions}(d), yielding only one path from $|000\rangle$ to 
$|000\rangle$ and the rescaled weight $w_{0,0}''$. The product of the edge 
weights for this path, $w_{0,0}''w_{1,1}'w_{2,2}$ is precisely the product of 
diagonal terms of $M$ in lower-triangular form, Eq.~(\ref{eq:w22pp}).

It is instructive to write the consequences of row reduction for the fermionic
representation of the eigenstate, Eq.~(\ref{eq:Hamfermionn=3}), for the example 
considered above. After the first round, the state becomes
\begin{eqnarray}
\tilde{M}_f^3|{\bf 0}\rangle
&=&w_{2,2}f_2^{\dag}\left(w_{1,0}'f_0^{\dag}+w_{1,1}'f_1^{\dag}\right)
\nonumber \\
&\times&\left(w_{0,0}'f_0^{\dag}+w_{0,1}'f_1^{\dag}\right)|{\mathcal O}\rangle,
\label{eq:round1fermions}
\end{eqnarray}
using the Pauli principle. After the second round, one obtains
\begin{eqnarray}
\tilde{M}_f^3|{\bf 0}\rangle&=&w_{2,2}f_2^{\dag}w_{1,1}'f_1^{\dag}
w_{0,0}''f_0^{\dag}|{\mathcal O}\rangle
=D_3f_2^{\dag}f_1^{\dag}f_0^{\dag}|{\mathcal O}\rangle.\hphantom{aa}
\end{eqnarray}
Thus, for fermions, no explicit expansion of the state~(\ref{eq:Hamfermionn=3}) 
is required; rather, the initial product of factors with three terms is reduced 
to a product of factors with two terms, and finally a product of single terms. 
The general strategy is the same for all $n$. This reduction of the evaluation 
of the determinant to a product of $n$ terms is at the heart of its efficiency.

\subsection{Example: Three Bosons}
\label{subsec:rowreducebosons}

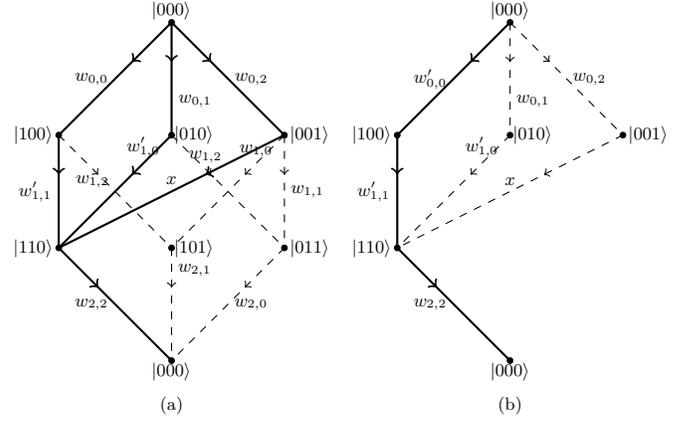
\begin{figure}[t]
\begin{tikzpicture}[decoration={
    markings,
    mark=at position 0.35 with {\arrow{>}}},
		scale=0.75,every node/.style={scale=0.75}
]


\draw[fill] (2.55, 2.0) arc(0:360:0.05) -- cycle;
\draw[fill] (0.55, 0.0) arc(0:360:0.05) -- cycle;
\draw[fill] (2.55, 0.0) arc(0:360:0.05) -- cycle;
\draw[fill] (4.55, 0.0) arc(0:360:0.05) -- cycle;
\draw[fill] (0.55, -2.0) arc(0:360:0.05) -- cycle;
\draw[fill] (2.55, -2.0) arc(0:360:0.05) -- cycle;
\draw[fill] (4.55, -2.0) arc(0:360:0.05) -- cycle;
\draw[fill] (2.55, -4.0) arc(0:360:0.05) -- cycle;

\draw[thick,black,postaction={decorate}] (2.5,2) -- (0.5,0)
  node [midway,left] {$w_{0,0}$};
\draw[thick,black,postaction={decorate}] (2.5,2) -- (2.5,0)
  node [pos=0.7,right] {$w_{0,1}$};
\draw[thick,black,postaction={decorate}] (2.5,2) -- (4.5,0)
  node [midway,right] {$w_{0,2}$};
\draw[thick,black,postaction={decorate}] (0.5,0) -- (0.5,-2)
  node [midway,left] {$w_{1,1}'$};
\draw[dashed,black,postaction={decorate}] (0.5,0) -- (2.5,-2)
  node [pos=0.3,below] {\small $w_{1,2}$};
\draw[thick,black,postaction={decorate}] (4.5,0) -- (0.5,-2)
  node [midway,above] {$x$};
\draw[thick,black,postaction={decorate}] (2.5,0) -- (0.5,-2)
  node [near start,above] {$w_{1,0}'$};
\draw[dashed,black,postaction={decorate}] (2.5,0) -- (4.5,-2)
  node [pos=0.3,above] {\small $w_{1,2}$};
\draw[dashed,black,postaction={decorate}] (4.5,0) -- (4.5,-2)
  node [midway,right] {\small $w_{1,1}$};
\draw[dashed,black,postaction={decorate}] (4.5,0) -- (2.5,-2)
  node [near start,above] {\small $w_{1,0}$};
\draw[thick,black,postaction={decorate}] (0.5,-2) -- (2.5,-4)
  node [midway,left] {$w_{2,2}$};
\draw[dashed,black,postaction={decorate}] (2.5,-2) -- (2.5,-4)
  node [pos=0.2,right] {\small $w_{2,1}$};
\draw[dashed,black,postaction={decorate}] (4.5,-2) -- (2.5,-4)
  node [midway,right] {\small $w_{2,0}$};

\draw[thick,black] (2.5,2.2) node {$|000\rangle$};
\draw[thick,black] (0.05,0.0) node {$|100\rangle$};
\draw[thick,black] (2.9,-0.0) node {$|010\rangle$};
\draw[thick,black] (4.95,0.0) node {$|001\rangle$};
\draw[thick,black] (0.05,-2.0) node {$|110\rangle$};
\draw[thick,black] (2.9,-2.0) node {$|101\rangle$};
\draw[thick,black] (4.95,-2.0) node {$|011\rangle$};
\draw[thick,black] (2.5,-4.2) node {$|000\rangle$};

\draw[] (2.5,-4.8) node {(a)};


\draw[fill] (8.55, 2.0) arc(0:360:0.05) -- cycle;
\draw[fill] (6.55, 0.0) arc(0:360:0.05) -- cycle;
\draw[fill] (8.55, 0.0) arc(0:360:0.05) -- cycle;
\draw[fill] (10.55, 0.0) arc(0:360:0.05) -- cycle;
\draw[fill] (6.55, -2.0) arc(0:360:0.05) -- cycle;
\draw[fill] (8.55, -4.0) arc(0:360:0.05) -- cycle;

\draw[thick,black,postaction={decorate}] (8.5,2) -- (6.5,0)
  node [midway,left] {$w_{0,0}'$};
\draw[dashed,black,postaction={decorate}] (8.5,2) -- (8.5,0)
  node [pos=0.7,right] {\small $w_{0,1}$};
\draw[dashed,black,postaction={decorate}] (8.5,2) -- (10.5,0)
  node [midway,right] {\small $w_{0,2}$};
\draw[thick,black,postaction={decorate}] (6.5,0) -- (6.5,-2)
  node [midway,left] {$w_{1,1}'$};
\draw[dashed,black,postaction={decorate}] (10.5,0) -- (6.5,-2)
  node [midway,above] {\small $x$};
\draw[dashed,black,postaction={decorate}] (8.5,0) -- (6.5,-2)
  node [near start,above] {\small $w_{1,0}'$};
\draw[thick,black,postaction={decorate}] (6.5,-2) -- (8.5,-4)
  node [midway,left] {$w_{2,2}$};

\draw[thick,black] (8.5,2.2) node {$|000\rangle$};
\draw[thick,black] (6.05,0.0) node {$|100\rangle$};
\draw[thick,black] (8.9,-0.0) node {$|010\rangle$};
\draw[thick,black] (10.95,0.0) node {$|001\rangle$};
\draw[thick,black] (6.05,-2.0) node {$|110\rangle$};
\draw[thick,black] (8.5,-4.2) node {$|000\rangle$};

\draw[] (8.5,-4.8) node {(b)};

\end{tikzpicture}
\caption{Round 1 (a) and 2 (b) of row reduction for the original spin model, 
equivalent to Schwinger bosons, for $n=3$.}
\label{fig:Hamn=3bosons}
\end{figure}

Consider next row reductions for the bosonic case, again using $n=3$ as an
example to illustrate the procedure for general $n$. The procedure works in
much the same way as for fermions, and is depicted in 
Fig.~\ref{fig:Hamn=3bosons}). The initial graph is equivalent to that for 
the original spin model, and is shown in Fig.~\ref{fig:Hamn=3}). 

The (unnormalized) rank-1 generalized zero eigenvectors of $\breve{M}_b$,
Eq.~(\ref{eq:Hamalt}), can be written as
\begin{eqnarray}
|v_1^{(1)}\rangle&=&|011\rangle-\frac{w_{2,0}}{w_{2,2}}|110\rangle;\nonumber \\
|v_2^{(1)}\rangle&=&|101\rangle-\frac{w_{2,1}}{w_{2,2}}|110\rangle.
\label{eq:v1n=3}
\end{eqnarray}
Comparison with Eq.~(\ref{eq:v1f}), one notices the similarity with 
$|v_2^{(1)}\rangle$ and $|v_3^{(1)}\rangle$ in the fermionic case, and also
that there is no rank-1 zero eigenvector involving $h=1$ states.
The vertices labeled by the bitstrings $011$ and $101$ can be eliminated
choosing the projector
\begin{eqnarray}
B_1&=&I-|v_1^{(1)}\rangle\langle 011|-|v_2^{(1)}\rangle\langle 101|\nonumber \\
&=&\begin{pmatrix}
1 & 0 & 0 & 0 & 0 & 0 & 0\cr
0 & 1 & 0 & 0 & 0 & 0 & 0\cr
0 & 0 & 1 & 0 & 0 & 0 & 0\cr
0 & 0 & 0 & 0 & 1 & 0 & 0\cr
0 & 0 & 0 & \frac{w_{2,0}}{w_{2,2}} & 0 & \frac{w_{2,1}}{w_{2,2}} & 1\cr
\end{pmatrix},
\end{eqnarray}
and
\begin{eqnarray}
A_1&=&\breve{M}_b\backslash \{\langle 011|,\langle 101|\}\nonumber \\
&=&\begin{pmatrix}
0 & 0 & 0 & 0 & w_{2,2}\cr
w_{0,2} & 0 & 0 & 0 & 0\cr
w_{0,1} & 0 & 0 & 0 & 0\cr
0 & w_{1,1} & w_{1,2} & 0 & 0\cr
w_{0,0} & 0 & 0  & 0 & 0\cr
0 & w_{1,0} & 0 & w_{1,2} & 0\cr
0 & 0 & w_{1,0} & w_{1,1} & 0\cr
\end{pmatrix}.
\end{eqnarray}
Again, it is straightforward to verify that $A_1B_1=\breve{M}$. One then obtains
\begin{equation}
\tilde{M}^{(2)}=B_1A_1=\begin{pmatrix}
0 & 0 & 0 & 0 & w_{2,2}\cr
w_{0,2} & 0 & 0 & 0 & 0\cr
w_{0,1} & 0 & 0 & 0 & 0\cr
w_{0,0} & 0 & 0 & 0 & 0\cr
0 & x & w_{1,0}' & w_{1,1}' & 0\cr
\end{pmatrix},
\end{equation}
where $w_{1,0}'$, and $w_{1,1}'$ coincide with the expressions in
Eq.~(\ref{eq:primed2}) but with minus signs replaced with plus signs; and a new
term is introduced,
\begin{equation}
x=\frac{w_{1,0}w_{2,1}+w_{1,1}w_{2,0}}{w_{2,2}}.
\end{equation}

The first round of row reductions, shown in Fig.~\ref{fig:Hamn=3bosons})(a), 
corresponds to deleting two vertices in the $h=2$ layer but none in the $h=1$
layer, in contrast with the fermionic case. Deleting vertices in only
a single layer avoids generating paths with rescaled weights on two adjacent 
edges, which would yield unphysical cross terms in their product (c.f.\ the 
discussion in Sec.~\ref{subsec:permderm}). But this comes at a high cost: 
vertices cannot be deleted in an adjacent layer simultaneously if they share
edges with vertices in the first layer, as is possible in the fermionic case.
This clearly decreases the efficiency of the reduction. Furthermore, 
deleting vertices in one layer requires adding new edges from the remaining 
vertex in that layer to all vertices in the next layer that had (now deleted) 
edges; in this case, the additional edge has weight $x$.

The second and final round of row reductions in this example is governed by the
rank-2 generalized zero eigenvectors:
\begin{eqnarray}
|v_1^{(2)}\rangle&=&|001\rangle-\frac{x}{w_{1,1}'}|100\rangle;\nonumber \\
|v_2^{(2)}\rangle&=&|010\rangle-\frac{w_{1,0}'}{w_{1,1}'}|100\rangle.
\label{eq:v2n=3}
\end{eqnarray}
As shown in Fig.~\ref{fig:Hamn=3bosons})(b), the vertices labeled by the 
bitstrings $001$ and $010$ are eliminated:
\begin{eqnarray}
B_2&=&I-|v_1^{(2)}\rangle\langle 001|-|v_2^{(2)}\rangle\langle 010|\nonumber \\
&=&\begin{pmatrix}
1 & 0 & 0 & 0 & 0\cr
0 & \frac{x}{w_{1,1}'} & \frac{w_{1,0}'}{w_{1,1}'} & 1 & 0\cr
0 & 0 & 0 & 0 & 1\cr
\end{pmatrix},
\end{eqnarray}
and
\begin{eqnarray}
A_2&=&\breve{M}^{(2)}\backslash \{\langle 001|,\langle 010|\}\nonumber \\
&=&\begin{pmatrix}
0 & 0 & w_{2,2}\cr
w_{0,2} & 0 & 0\cr
w_{0,1} & 0 & 0\cr
w_{0,0} & 0 & 0\cr
0 & w_{1,1}' & 0\cr
\end{pmatrix}.
\end{eqnarray}
One then obtains
\begin{equation}
\tilde{M}^{(3)}=B_2A_2=\begin{pmatrix}
0 & 0 & w_{2,2}\cr
w_{0,0}' & 0 & 0\cr
0 & w_{1,1}' & 0\cr
\end{pmatrix},
\end{equation}
where
\begin{equation}
w_{0,0}'=w_{0,0}+\frac{w_{0,1}w_{1,0}'+w_{0,2}x}{w_{1,1}'}.
\end{equation}
As is shown in Fig.~\ref{fig:Hamn=3bosons}(b), two vertices in the $h=1$ layer 
are now deleted, requiring a rescaling of the $w_{0,0}$ weight, and one obtains 
a single path from $|000\rangle$ to $|000\rangle$, as desired.
The permanent is then
\begin{equation}
P_3=w_{0,0}'w_{1,1}'w_{2,2},
\end{equation}
which is expressed as a product of three single terms, much like the expression
of the determinant in Eq.~(\ref{eq:detn=3}).

\subsection{Example: Four Bosons}
\label{subsec:rowreduce4bosons}

Given the superficial similarities between row reductions for the bosonic and
fermionic cases when $n=3$, it is instructive to discuss the $n=4$ case to 
gather a better understanding of why the permanent is nevertheless 
exponentially more difficult to compute with this method. 
The rank-1 generalized zero eigenvectors are
\begin{eqnarray}
|v_1^{(1)}\rangle&=&|0011\rangle-\frac{w_{2,1}}{w_{2,3}}|0110\rangle
-\frac{w_{2,0}}{w_{2,2}}|1001\rangle\nonumber
+\frac{w_{2,0}w_{2,1}}{w_{2,2}w_{2,3}}\nonumber \\
&=&\left(|01\rangle-\frac{w_{2,0}}{w_{2,2}}|10\rangle\right)_{0,2}
\left(|01\rangle-\frac{w_{2,1}}{w_{2,3}}|10\rangle\right)_{1,3};\nonumber \\
|v_2^{(1)}\rangle&=&|0101\rangle-\frac{w_{2,2}}{w_{2,3}}|0110\rangle
-\frac{w_{2,0}}{w_{2,1}}|1001\rangle
+\frac{w_{2,0}w_{2,2}}{w_{2,1}w_{2,3}};\nonumber \\
&=&\left(|01\rangle-\frac{w_{2,0}}{w_{2,1}}|10\rangle\right)_{0,1}
\left(|01\rangle-\frac{w_{2,2}}{w_{2,3}}|10\rangle\right)_{2,3};\nonumber \\
|v_3^{(1)}\rangle&=&|0111\rangle-\frac{w_{3,0}}{w_{3,3}}|1110\rangle\nonumber \\
&=&\left(|01\rangle-\frac{w_{3,0}}{w_{3,3}}|10\rangle\right)_{0,3}
|11\rangle_{1,2};\nonumber \\
|v_4^{(1)}\rangle&=&|1011\rangle-\frac{w_{3,1}}{w_{3,3}}|1110\rangle\nonumber
 \\
&=&\left(|01\rangle-\frac{w_{3,1}}{w_{3,3}}|10\rangle\right)_{1,3}
|11\rangle_{0,2};\nonumber \\
|v_5^{(1)}\rangle&=&|1101\rangle-\frac{w_{3,2}}{w_{3,3}}|1110\rangle\nonumber \\
&=&\left(|01\rangle-\frac{w_{3,2}}{w_{3,3}}|10\rangle\right)_{2,3}
|11\rangle_{0,1}.
\label{eq:v1n=4}
\end{eqnarray}
The eigenvectors can all be written in explicitly separable forms, where the
indices outside the parentheses denotes the label partitions; note that these
also match the second indices in the weight ratios. Evidently the 
the generalized zero eigenvectors for $n=3$, Eqs.~(\ref{eq:v1n=3}) and 
(\ref{eq:v2n=3}), can be written in a similar product form. This is due to the 
fact that the Hamiltonian~(\ref{eq:Hamalt}) itself can be written as a sum of 
permutations of separable terms. For example, the terms in the $n=4$ 
Hamiltonian that map $h=2$ states to $h=3$ states can be expressed as
\begin{eqnarray}
\breve{M}_{(h=2\to 3)}&=&\frac{1}{2}\Big[()_{0,1}I_{2,3}
+I_{0,1}()_{2,3}+()_{0,2}I_{1,3}+I_{0,2}()_{1,3}\nonumber \\
&+&()_{0,3}I_{1,2}+I_{0,3}()_{1,2}\Big],
\label{eq:Mh2to3}
\end{eqnarray}
where 
\begin{eqnarray}
()_{i,j}&=&|11\rangle_{i,j}\left(w_{2,i}\langle 01|+w_{2,j}\langle 10|
\right)_{i,j},
\nonumber \\
I_{i,j}&=&\left(|01\rangle\langle 01|+|10\rangle\langle 10|\right)_{i,j}.
\end{eqnarray}
The `identity' operator is the sum of all idempotents with $h=1$, enumerated
by the $1/2$ prefactor in Eq.~(\ref{eq:Mh2to3}). Thus, $\breve{M}_{(h=2\to 3)}$ 
has the form of Cartesian products of operators over all four-site bipartitions 
restricted to states with specific Hamming weight. It is straightforward to 
verify that the $|v_1^{(1)}\rangle$ and $|v_2^{(1)}\rangle$ in 
Eq.~(\ref{eq:v1n=4}) are zero eigenvectors of the first and second Cartesian 
product, respectively, and have no support on the third. Similar expressions
can be obtained for the other terms in the Hamiltonian.

Construction of $A_1$ and $B_1$ proceeds analogously to the $n=3$ case, and 
generates $\breve{M}^{(2)}=B_1A_1$ with basis states (graph vertices) 
$\{|0011\rangle,|0101\rangle,|0111\rangle,|1011\rangle,|1101\rangle\}$ removed. 
However, all but one of the 23 non-zero terms in the resulting matrix is 
unique. From the graph perspective, only 5 edges are reweighted, 9 edges have 
unchanged weights, and 9 new edges with unique weights must be added. 
Generically, in the bosonic case, the number of unique terms that need to be 
evaluated during the row reduction procedure grows exponentially with $n$.
There doesn't appear to be any way to exploit the separable nature of the 
generalized zero eigenvectors to simplify the calculation.

\section{Dicussion: Prospects for a quantum algorithm}
\label{sec:prospects}

As discussed in Sec.~\ref{subsec:eigensystem}, the permanent of an $n\times n$ 
matrix $M$ relates to the eigenvalues of another $2^n$-dimensional matrix
$\tilde{M}$ or $\breve{M}$ (the dimension of the latter is one smaller so that
one basis state is unused). This matrix has several attributes that would 
appear to favor the development of an efficient quantum algorithm for the 
evaluation of the permanent: the dimension of $\tilde{M}$ is a power of two, 
which would be the case for an $n$-qubit operator; matrix elements of 
$\tilde{M}$ are easily indexed by address, which correspond to their original 
positions in $M$; $\tilde{M}$ is $n$-sparse (no row or column has more than 
$n-1$ elements); and the permanent is the maximal eigenvalue of $\tilde{M}^n$. 
Despite these nice features, however, the construction of an efficient 
algorithm for the permanent using this approach is not straightforward for one 
principal reason: neither $\tilde{M}$ nor $\breve{M}$ is Hermitian or unitary.

As a first attempt at a quantum algorithm, one might leverage the relation 
$P_n=\braket{0|\breve{M}^n|0}$, Eq.~\eqref{eq:easyperm}. The quantity on the 
right-hand side can be computed using any of the known algorithms for 
evaluating expectation values~\cite{Knill2007,Rall2020}. Unfortunately, such 
algorithms have $O(1/\epsilon)$ or worse dependence on additive 
error~\cite{Alase2022}, and thus an even worse dependence on the multiplicative 
error. Moreover, the operator norm of $\breve{M}^n$ is not polynomially bounded 
in general. Consequently, this approach fails to suggest an avenue toward an 
efficient quantum algorithm.

A more lucrative approach might be to make use of the fact that all non-zero 
eigenvalues of $\breve{M}$ have absolute value $|P_n|^{1/n}$, 
Eq.~\eqref{eq:eigenvalues}. Thus, if there exists an efficient procedure to
generate one of the corresponding eigenstates, then $|P_n|^{1/n}$ can be 
computed efficiently to constant or polynomially small additive error. Note 
that an additive approximation of $|P_n|^{1/n}$ provides significantly more 
resolution, at least for the unitary matrices $M$ that would be relevant to
boson sampling, in contrast to an additive approximation of $|P_n|$. As noted 
by Aaronson and Arkhipov~\cite{Aaronson2011b}, since $|P_n|$ is typically 
exponentially small 
for unitary matrices sampled from a Haar random distribution, an additive 
approximation of $|P_n|$ to polynomial accuracy would almost always return 
zero. On the other hand, the average of $|P_n|^{1/n}$ is in $\Omega(1)$ for unitary 
matrices sampled from a Haar random distribution, and therefore an 
approximation of $|P_n|^{1/n}$ to polynomially small or even constant additive 
error provides more resolution. 

Unfortunately, generating any eigenstates of $\breve{M}$, 
Eq.~(\ref{eq:eigenvectors}), is not straightforward. The coefficients in the 
linear combination depend on the value of $P_n$, which is unknown and in fact 
the goal of the computation. Even if one could obtain a sufficiently good 
approximation of $P_n$ (via randomized classical algorithms), taking 
appropriate linear combinations using the techniques developed in 
Ref.~\cite{Berry2015} would only generate the targeted eigenstate with 
exponentially small probability, limiting the runtime of the algorithm. 
Specifically, it is not obvious how to adapt block-encoding 
techniques~\cite{Low2019} to generate, with high probability, the state 
$\breve{M}^{n-1}|0\rangle/\|\breve{M}^{n-1}|0\rangle\|$, which is one of the 
terms in the desired linear combination.
    
Consider instead leveraging the useful property that the state $\ket{0}$ is an 
equal superposition of all eigenstates corresponding to non-zero eigenvalues of 
$\breve{M}$. If $\breve{M}$ were unitary, this fact would have been sufficient 
for computing all eigenvalues of $\breve{M}$ efficiently using repeated 
application of the phase estimation algorithm~\cite{Abrams1999}. Unfortunately, 
the extension of phase estimation to non-unitary operators is generally 
inefficient~\cite{Wang}. For the present problem, we expect phase estimation 
to take an exponentially long time, as the eigenvalues being estimated lie well
inside the unit circle. 

A more sophisticated approach to computing the eigenvalues of $\breve{M}$ is
based on quantum linear-system solvers~\cite{Shao2022}. The complexity of this 
approach is limited by the condition number of the eigenvectors, however. We 
have verified numerically that the condition number is exponentially large for 
typical real and complex matrices $M$. 

To summarize, mapping the problem of computing the permanent of $M$ to 
calculating the eigenvalues of $\breve{M}$ would seem to suggest new routes for 
designing an efficient quantum algorithm to obtain a multiplicative 
approximation of the permanent. Yet, such an algorithm does not follow from the
immediate application of the currently available algorithmic tools for linear 
algebra in a quantum setting. In all likelihood, if such an algorithm exists,
it would rely on more subtle properties of the permanent than are made apparent 
by the present mapping.

\begin{widetext}
\appendix
\section{Block-diagonal representation}
\label{app:block}

This section derives the expression for $\tilde{M}^{n+1}$, where $\tilde{M}$ is 
defined by Eq.~(\ref{eq:Ham}). Consider first 
$\tilde{M}^2$:
\begin{eqnarray}
\tilde{M}^2&=&\left(\sum_{{\bf i}_0}\sum_{j_0=0}^{n-1}w_{h({\bf i}_0),j_0}
\sigma^+_{j_0}|{\bf i}_0\rangle\langle{\bf i}_0|+|{\bf 0}\rangle\langle{\bf 1}|
\right)\left(\sum_{{\bf i}_1}\sum_{j_1=0}^{n-1}w_{h({\bf i}_1),j_1}
\sigma^+_{j_1}|{\bf i}_1\rangle\langle{\bf i}_1|+|{\bf 0}\rangle\langle{\bf 1}|
\right) \nonumber \\
&=&\sum_{{\bf i}_0,{\bf i}_1}\sum_{j_0,j_1}w_{h({\bf i}_0),j_0}\sigma^+_{j_0}
|{\bf i}_0\rangle\langle{\bf i}_0|w_{h({\bf i}_1),j_1}
\sigma^+_{j_1}|{\bf i}_1\rangle\langle{\bf i}_1|
+\sum_{{\bf i}_1}\sum_{j_1}|{\bf 0}\rangle\langle{\bf 1}|w_{h({\bf i}_1),j_1}
\sigma^+_{j_1}|{\bf i}_1\rangle\langle{\bf i}_1|
+\sum_{{\bf i}_0}\sum_{j_0}w_{h({\bf i}_0),j_0}\sigma^+_{j_0}
|{\bf i}_0\rangle\langle{\bf i}_0|{\bf 0}\rangle\langle{\bf 1}|.\nonumber \\
\label{eq:M2}
\end{eqnarray}
For the evaluation of the $\langle{\bf i}_0|w_{h({\bf i}_1),j_1}
\sigma^+_{j_1}|{\bf i}_1\rangle$ factor in first term of Eq.~(\ref{eq:M2}), 
there are three
possibilities: $\sigma^+_{j_1}|{\bf i}_1\rangle=0$, $\sigma^-_{j_1}|{\bf i}_0
\rangle=0$, or $\sigma^+_{j_1}|{\bf i}_1\rangle=|{\bf i}_0\rangle$ 
(equivalently $|{\bf i}_1\rangle=\sigma^-_{j_1}|{\bf i}_0\rangle$ or 
$\langle{\bf i}_1|=\langle{\bf i}_0|\sigma^+_{j_1}$), and the first two 
possibilities contribute nothing to the sum. Similar arguments apply to the
second and third terms, and one obtains
\begin{eqnarray}
\tilde{M}^2&=&
\sum_{{\bf i}}\sum_{j_0,j_1}w_{h({\bf i}),j_0}w_{h({\bf i})-1,j_1}
\sigma^+_{j_0}|{\bf i}\rangle\langle{\bf i}|\sigma_{j_1}^+
+\sum_{j_0}\left(w_{n-1,j_0}|{\bf 0}\rangle\langle{\bf 1}|\sigma_{j_1}^+
+w_{0,j_0}\sigma^+_{j_0}|{\bf 0}\rangle\langle{\bf 1}|\right).
\label{eq:Mn2}
\end{eqnarray}
Next consider $\tilde{M}^3$. After elementary algebra along the same lines as
above, one obtains
\begin{eqnarray}
\tilde{M}^3&=&\sum_{{\bf i}}\sum_{j_0,j_1,j_2}\left(w_{h({\bf i}),j_0}
\sigma^+_{j_0}\right)|{\bf i}\rangle\langle{\bf i}|
\left(w_{h({\bf i})-1,j_1}\sigma_{j_1}^+\right)
\left(w_{h({\bf i})-2,j_2}\sigma_{j_2}^+\right)
+\sum_{j_0,j_1}\left(w_{0,j_0}\sigma^+_{j_0}\right)\left(
w_{1,j_1}\sigma^+_{j_1}\right)|{\bf 0}\rangle\langle{\bf 1}|\nonumber \\
&+&\sum_{j_{n-1},j_{n-2}}|{\bf 0}\rangle\langle{\bf 1}|
\left(w_{n-1,j_{n-1}}\sigma_{j_{n-1}}^+\right)\left(w_{n-2,j_{n-2}}
\sigma_{j_{n-2}}^+\right)
+\sum_{j_0,j_{n-1}}\left(w_{0,j_0}\sigma^+_{j_0}\right)|{\bf 0}\rangle
\langle{\bf 1}|\left(w_{n-1,j_{n-1}}\sigma_{j_{n-1}}^+\right).
\label{eq:Mn3}
\end{eqnarray}
The form of leading term in $\tilde{M}^{n+1}$ should now be evident:
\begin{eqnarray}
&&\sum_{{\bf i}}\sum_{j_0,\ldots,j_n}\left(w_{h({\bf i}),j_0}
\sigma^+_{j_0}\right)|{\bf i}\rangle\langle{\bf i}|\left(w_{h({\bf i})-1,j_1}
\sigma_{j_1}^+\right)\cdots \left(w_{h({\bf i})-n,j_n}\sigma_{j_n}^+\right).
\end{eqnarray}
In the above expression, the $\langle{\bf i}|\prod_k\sigma_{j_k}^+$ term is 
zero unless ${\bf i}={\bf 1}$, but then
$\sigma^+_{j_0}|{\bf i}\rangle=0$, so that the leading term vanishes. The 
remaining terms are straightforward generalizations of those found in 
Eqs.~(\ref{eq:Mn2}) and (\ref{eq:Mn3}), and one obtains
\begin{eqnarray}
\tilde{M}^{n+1}&=&\sum_{j_0,\ldots,j_{n-1}}\left[
\left(w_{0,j_0}\sigma^+_{j_0}\right)\cdots\left(w_{n-1,j_{n-1}}
\sigma^+_{j_{n-1}}\right)|{\bf 0}\rangle\langle{\bf 1}|
+\left(w_{0,j_0}\sigma^+_{j_0}\right)\cdots\left(w_{n-2,j_{n-2}}
\sigma^+_{j_{n-2}}\right)|{\bf 0}\rangle\langle{\bf 1}|
\left(w_{n-1,j_{n-1}}\sigma^+_{j_{n-1}}\right)\right.\nonumber \\
&+&\ldots+\left.\left(w_{0,j_0}\sigma^+_{j_0}\right)|{\bf 0}\rangle
\langle{\bf 1}|\left(w_{1,j_1}\sigma^+_{j_1}\right)\cdots\left(w_{n-1,j_{n-1}}
\sigma^+_{j_{n-1}}\right)+|{\bf 0}\rangle\langle{\bf 1}|\left(w_{0,j_0}
\sigma^+_{j_0}\right)\cdots\left(w_{n-1,j_{n-1}}\sigma^+_{j_{n-1}}\right)
\right].
\label{eq:Mnapp}
\end{eqnarray}

A corollary is that the expression for arbitrary powers $p$ is
\begin{eqnarray}
\tilde{M}^p&=&\sum_{{\bf i}}\sum_{j_0,\ldots,j_{p-1}}\left(w_{h({\bf i}),j_0}
\sigma^+_{j_0}\right)|{\bf i}\rangle\langle{\bf i}|
\left(w_{h({\bf i})-1,j_1}\sigma_{j_1}^+\right)\cdots
\left(w_{h({\bf i})-p+1,j_{p-1}}\sigma_{j_{p-1}}^+\right)\nonumber \\
&+&\sum_{j_0,\ldots,j_{p-2}}\left[
\left(w_{0,j_0}\sigma^+_{j_0}\right)\cdots\left(w_{p-2,j_{p-2}}
\sigma^+_{j_{p-1}}\right)|{\bf 0}\rangle\langle{\bf 1}|
+\left(w_{0,j_0}\sigma^+_{j_0}\right)\cdots\left(w_{p-3,j_{p-3}}
\sigma^+_{j_{n-2}}\right)|{\bf 0}\rangle\langle{\bf 1}|
\left(w_{p-2,j_{p-2}}\sigma^+_{j_{p-2}}\right)\right.\nonumber \\
&+&\ldots+\left.
|{\bf 0}\rangle\langle{\bf 1}|\left(w_{n-1,j_0}
\sigma^+_{j_0}\right)\cdots\left(w_{n-p+1,j_{p-2}}\sigma^+_{j_{p-2}}\right)
\right],
\label{eq:Mpapp}
\end{eqnarray}
which can be used to prove Eq.~(\ref{eq:Mblock}), i.e.\ that 
\begin{equation}
\tilde{M}_m=\tilde{M}^m|{\bf 0}\rangle\langle{\bf 0}|\tilde{M}^{n-m+1}.
\end{equation}
First,
\begin{equation}
\tilde{M}^p|{\bf 0}\rangle=\sum_{{\bf i}}\sum_{j_0,\ldots,j_{p-1}}
\left(w_{h({\bf i}),j_0}
\sigma^+_{j_0}\right)|{\bf i}\rangle\langle{\bf i}|
\left(w_{h({\bf i})-1,j_1}\sigma_{j_1}^+\right)\cdots
\left(w_{h({\bf i})-p+1,j_{p-1}}\sigma_{j_{p-1}}^+\right)|{\bf 0}\rangle.
\end{equation}
Only bitstrings ${\bf i}$ with Hamming weight $p-1$ will contribute, so
\begin{equation}
\tilde{M}^p|{\bf 0}\rangle=\sum_{j_0,\ldots,j_{p-1}}
\left(w_{0,j_0}\sigma^+_{j_0}\right)
\left(w_{1,j_1}\sigma_{j_1}^+\right)\cdots
\left(w_{p-1,j_{p-1}}\sigma_{j_{p-1}}^+\right)|{\bf 0}\rangle.
\end{equation}
Second, following similar reasoning,
\begin{eqnarray}
\langle{\bf 0}|\tilde{M}^q&=&\sum_{{\bf i}}\sum_{j_0,\ldots,j_{q-1}}
\langle{\bf 0}|w_{h({\bf i}),j_0}\sigma^+_{j_0}|{\bf i}\rangle\langle{\bf i}|
\left(w_{h({\bf i})-1,j_1}\sigma_{j_1}^+\right)\cdots
\left(w_{h({\bf i})-q+1,j_{q-1}}\sigma_{j_{q-1}}^+\right)\nonumber \\
&+&\sum_{j_0,\ldots,j_{q-2}}\langle{\bf 1}|\left(w_{n-1,j_0}
\sigma^+_{j_0}\right)\cdots\left(w_{n-q+1,j_{q-2}}\sigma^+_{j_{q-2}}\right)
\nonumber \\
&=&\sum_{j_0,\ldots,j_{q-2}}\langle{\bf 1}|\left(w_{n-1,j_0}
\sigma^+_{j_0}\right)\cdots\left(w_{n-q+1,j_{q-2}}\sigma^+_{j_{q-2}}\right).
\end{eqnarray}
Putting these results together:
\begin{eqnarray}
\tilde{M}^m|{\bf 0}\rangle\langle{\bf 0}|\tilde{M}^{n-m+1}
&=&\sum_{{j_0,\ldots,j_{m-1}\atop k_0,\ldots,k_{n-m-1}}}
\left(w_{0,j_0}\sigma^+_{j_0}\right)
\left(w_{1,j_1}\sigma_{j_1}^+\right)\cdots
\left(w_{m-1,j_{m-1}}\sigma_{j_{m-1}}^+\right)|{\bf 0}\rangle\nonumber \\
&&\qquad\times\langle{\bf 1}|\left(w_{n-1,k_0}\sigma^+_{k_0}\right)
\left(w_{n-2,k_1}\sigma^+_{k_1}\right)\cdots\left(w_{m,k_{n-m-1}}
\sigma^+_{k_{n-m-1}}\right)\nonumber \\
&=&\sum_{j_0,\ldots,j_{n-1}}\left(w_{0,j_0}\sigma^+_{j_0}\right)
\left(w_{1,j_1}\sigma_{j_1}^+\right)\cdots
\left(w_{m-1,j_{m-1}}\sigma_{j_{m-1}}^+\right)|{\bf 0}\rangle\nonumber \\
&&\qquad\times\langle{\bf 1}|\left(w_{n-1,j_m}\sigma^+_{j_m}\right)
\left(w_{n-2,j_{m+1}}\sigma^+_{j_{m+1}}\right)\cdots\left(w_{m,j_{n-1}}
\sigma^+_{j_{n-1}}\right).
\end{eqnarray}
Comparison with the terms in Eq.~(\ref{eq:Mn}) immediately yields
\begin{equation}
\tilde{M}_m=\tilde{M}^m|{\bf 0}\rangle\langle{\bf 0}|\tilde{M}^{n-m+1}.
\label{app:Mm}
\end{equation}

\end{widetext}

\bibliographystyle{apsrev.bst}
\bibliography{ref3}
\end{document}